\DocumentMetadata{}
\documentclass[acmsmall]{acmart}

\usepackage{xcolor}
\usepackage{multicol}
\usepackage{subcaption}
\usepackage{xcolor}
\usepackage[normalem]{ulem}
\useunder{\uline}{\ul}{}
\usepackage{xcolor}

\useunder{\uline}{\ul}{}

\usepackage{multirow}
\usepackage{tablefootnote}

\usepackage{lipsum}
\usepackage{tcolorbox}
\usepackage{etoolbox}
\usepackage{setspace} 
\usepackage{tabularx}

\usepackage{longtable}
\usepackage{array}
\definecolor{blue}{RGB}{65,105,225}
\definecolor{green}{RGB}{0, 100, 0} 
\definecolor{bluec}{RGB}{0,0,0}
\definecolor{orangec}{RGB}{255, 165, 0}
\definecolor{azure}{rgb}{0.0, 0.5, 1.0}

\definecolor{ao}{rgb}{0.0, 0.0, 1.0}

\AtBeginEnvironment{quote}{\par\singlespacing\small}

\definecolor{neonfuchsia}{rgb}{1.0, 0.25, 0.39}
\newcommand{\yiren}[1]{{\small\textcolor{neonfuchsia}{\bf [*** Yi-Ren: #1]}}}
\newcommand{\yirenrevised}[1]{\textcolor{bluec}{#1}}

\newcommand{\yun}[1]{\textcolor{ao}{[\textbf{yun}: #1]}}

\AtBeginDocument{%
  \providecommand\BibTeX{{%
    \normalfont B\kern-0.5em{\scshape i\kern-0.25em b}\kern-0.8em\TeX}}}

\acmYear{2024}

\begin{document}

\title{\yirenrevised{Improving Emotional Support Delivery in Text-Based Community Safety Reporting Using Large Language Models}}

\author{Yiren Liu}
\email{yirenl2@illinois.edu}
\affiliation{%
  \institution{Informatics, University of Illinois Urbana-Champaign}
  \country{USA}
}

\author{Yerong Li}
\email{yerong2@illinois.edu}
\affiliation{%
  \institution{Department of Computer Science, University of Illinois Urbana-Champaign}
  \country{USA}
}

\author{Ryan Mayfield}
\email{ryan@livesafemobile.com}
\affiliation{%
  \institution{LiveSafe Inc.}
  \country{USA}
}

\author{Yun Huang}
\email{yunhuang@illinois.edu}
\affiliation{%
  \institution{School of Information Sciences, University of Illinois Urbana-Champaign}
  \country{USA}
}

\begin{abstract}

\yirenrevised{Emotional support is a crucial aspect of communication between community members and police dispatchers during incident reporting. However, there is a lack of understanding about how emotional support is delivered through text-based systems, especially in various non-emergency contexts. In this study, we analyzed two years of chat logs comprising 57,114 messages across 8,239 incidents from 130 higher education institutions. Our empirical findings revealed significant variations in emotional support provided by dispatchers, influenced by the type of incident, service time, and a noticeable decline in support over time across multiple organizations. To improve the consistency and quality of emotional support, we developed and implemented a fine-tuned Large Language Model (LLM), named \textit{dispatcherLLM}. We evaluated \textit{dispatcherLLM} by comparing its generated responses to those of human dispatchers and other off-the-shelf models using real chat messages. Additionally, we conducted a human evaluation to assess the perceived effectiveness of the support provided by \textit{dispatcherLLM}. This study not only contributes new empirical understandings of emotional support in text-based dispatch systems but also demonstrates the significant potential of generative AI in improving service delivery.}

\end{abstract}



\ccsdesc[500]{Human-centered computing~Empirical studies in collaborative and social computing}

\keywords{text-based reporting system, live chat, safety reporting, emotion classification, large language model}

\maketitle

\section{Introduction}
\yirenrevised{The importance of emotional support in public safety services is well-documented, particularly in voice-based emergency response systems \cite{Observations_on_the_Display, feldman2021calming, paoletti2012operators, tracy2002questioning}. Studies show that delivering emotional support effectively through communication is crucial for rapidly establishing rapport and trust, which facilitates quicker and more accurate information exchange~\cite{paoletti2012operators}. Such an empathetic approach is adopted in various emergency response practices to build trust \cite{Observations_on_the_Display} and reduce the initial distress of individuals affected by traumatic experiences \cite{sijbrandij2020effect}. These insights are aimed at enhancing service quality and improving the provision of emotional support within the safety reporting domain \cite{tracy1998emotion}.}

\yirenrevised{As community Information and Communication Technology (ICTs) have evolved, text-based safety reporting systems~\cite{ming2021examining,ignaco2019development,Ahmed:2014:PPF:2611205.2557376,doi:10.1080/1369118X.2017.1309444} are increasingly used by organizations for risk management. For instance, many universities and public organizations offer mobile applications that allow their community members to submit text reports about potential risks, concerning situations, or suspicious activities~\cite{ignaco2019development,Ahmed:2014:PPF:2611205.2557376,doi:10.1080/1369118X.2017.1309444}. Additionally, city police agencies have implemented text-to-911 that enable live chat with dispatchers~\cite{grace2021text,Textto911WhatYouNeedToKnow}.
Recent works \cite{ming2021examining, liu2023discovering} have demonstrated the effectiveness of text-based reporting systems, such as reducing the labor costs associated with safety service provision, enhancing user-perceived value, and fostering inclusivity by assisting marginalized groups in reporting risks.
Despite their growing popularity and demonstrated benefits, there remains a significant gap in understanding the emotional dynamics and support delivery within these text-based reporting systems.}


\yirenrevised{To address this gap, we conducted a comprehensive system log analysis of a text-based community safety reporting system, named LiveSafe \cite{ming2021examining}, which serves over 200 higher education institutions. This analysis helped us identify patterns of emotional expressions and support delivered in text-based incident reporting. We also observed a longitudinal decline in the provision of emotional support across various organizations, which prompted the development of new technical solutions. Although recent studies have demonstrated the promise and significant implications of using AI and Large Language Models (LLMs) to provide emotional support in various application domains \cite{chung2023challenges, zheng2023building, li2021towards,jo2023understanding}, little is known about whether AI models could be trained to support dispatchers in providing emotional support. Therefore, building on the empirical findings and leveraging the real chat logs, we developed an LLM, called \textit{dispatcherLLM}, fine-tuned for the domain of text-based safety incident reporting.
}

\yirenrevised{In the remainder of this paper, we refer to the community members who submitted reports as ``users'' and the human agents from the safety organizations who handled the reports via text as ``dispatchers.'' Our study makes novel and significant contributions to the CSCW and HCI communities: }
\yirenrevised{
\begin{itemize}
    \item First, we performed an emotion analysis on real text-based incident reporting systems. New empirical findings showed that users' emotions tend to be more negative when reporting specific categories of incidents, such as those related to \textit{Mental Health}. However, their messages generally become more positive through interactions with dispatchers.  
    \item Second, we found that the delivery of emotional support by dispatchers was associated with factors related to the incident, service timing, and organizational characteristics.
    A surprising finding was that organizations with prolonged use of the system tended to provide less emotional support. Notably, emotional support tends to diminish during certain working hours of the day (e.g., 8 a.m. - 12 p.m. and 12 p.m. - 4 p.m.). 
    \item Additionally, we developed and implemented an LLM fine-tuned using domain-specific chat logs to mitigate the two issues mentioned above. Our evaluation, using both automatic metrics and user surveys, demonstrated that our \textit{dispatcherLLM} could provide more consistent and effective emotional support compared to both human dispatchers and other general LLMs (not trained with domain-specific chats) for various incidents. 
    \item Lastly, our findings offer strong empirical and practical value and shed light on human-AI co-creation of public services~\cite{zheng2023synergizing}. Specifically, this research pioneers the use of LLM to improve emotional support in public safety services, while LLMs are being increasingly utilized across other service domains \cite{gelbrich2021emotional, ali2023supporting, meng2023mediated} to address the labor shortage. Given the sensitive nature of public safety, our goal is not to replace human workers with AI-based solutions. Instead, it is crucial to explore how LLMs can be leveraged to provide more consistent responses. 
We discuss several implications for the future design of text-based risk reporting systems, aiming to enhance service quality and alleviate dispatcher burnout.
\end{itemize}
}

\section{Related Work}
In this section, we present prior work on the importance and value of emotional support in safety incident reporting service. We then discuss existing systems and studies designing text-based risk reporting systems. 
Finally, we discuss recent works on AI-enabled systems providing emotional support.

\subsection{The Role of Emotional Support in Existing Safety Incident Reporting Systems} 
In emergency response systems, the delivery of emotional support is critical, particularly due to the ``emotional pain'' experienced by individuals reporting traumatic or safety-related incidents \cite{Observations_on_the_Display}. 
This support, when delivered effectively by dispatchers, is crucial for successful information gathering, enhancing trust, and fostering cooperation from callers.

The benefits of such proficient emotional support are significant, contributing to the recovery of individuals from traumatic experiences \cite{sijbrandij2020effect}. 
Meanwhile, inadequate emotional support can negatively impact the dispatcher-caller relationship. Inappropriate questioning strategies, for instance, have been perceived as delaying assistance, leading to user frustration and reluctance in cooperation \cite{paoletti2012operators,feldman2021calming}. 
This is further supported by findings that demonstrate that a lack of emotional support can result in callers perceiving questions as ``face-threatening,'' thereby resisting responding to questions \cite{tracy2002questioning}. 
This body of research highlights the crucial role of emotional support in incident handling systems, not only in facilitating immediate information collection but also in establishing a positive, cooperative rapport between dispatchers and callers for effective emergency response.

However, there is currently a lack of examination regarding the delivery of emotional support within text-based reporting systems and its potential impact on the service delivery of safety organizations. Existing work only discussed the functional aspects of the conversations between users and safety administrative agents using text-based reporting systems \cite{liu2023discovering}, e.g., the response rates (how many of the reports were handled) and responsiveness (how quickly they were handled). 
The issues reported by users via mobile or web-based applications are frequently less urgent than those reported through the 911 emergency hotline, as evidenced in the study by \citet{iriberri2006reporting}. 
These user-generated reports, also referred to as ``tips'' later in this study, typically detail less critical situations compared to the dire emergencies usually reported through the traditional 911 system.
Prior studies also showed that a person's speech and voice could convey more information regarding the person's mental state and emotional experience \cite{pinker1990natural, scherer2001emotion, hall2007sources}. 
Although the emotional aspect of incident reporting has been well discussed by prior studies, the previous accounts are limited to the context of emergency call-taking but not text-based reporting systems. 
Meanwhile, there has been a lack of empirical understanding about how people show their emotional state in text-based incident reporting and how service providers respond to different emotional states by delivering their service.

\subsection{Community Risk Reporting Systems}

Information and Communication Technology (ICT) has been increasingly utilized in risk management in public settings \cite{sivvcevic2020possibilities}. Several systems have been developed to enable the reporting of safety events and incidents by the community, including web-based forms, mobile apps, and systems that allow users to report safety concerns and risks through crowd-sourcing \cite{iriberri2006reporting, sakpere2015usable, blom2010fear, cvijikj2015towards}. Some systems also connect users with emergency service providers and automatically share their location, improving the effectiveness of getting assistance \cite{hossain2018bangladesh, kalyanchakravarthy2014android, ahmed2014protibadi}. In addition, live-chat safety incident reporting features have been incorporated into risk management systems by various organizations \cite{uicSAFE, ming2021examining, wetip}. Previous research has found that campus community safety apps with emergency texting features have a higher perceived utility by community members and a lower cost compared to traditional emergency communication systems such as blue-light emergency phone towers \cite{reaves2015campus, yabe2017cost}. Text-based reporting systems, such as text-to-911, have also been found to be beneficial for deaf and hard-of-hearing individuals and people with disabilities \cite{Textto911WhatYouNeedToKnow}.

Existing work has also suggested that how technological systems are designed can greatly impact users' perception of the target tasks and even gradually shift their behaviors. 
The implementation and adaptation of newly introduced technological systems into an existing organization workflow have been a non-trivial problem widely discussed by existing studies. \citet{tyre1994windows} revealed the highly discontinuous nature of technological adaptation in organizations, which usually happens within a relatively brief window after the initial implementation of a new technology. 
A later study by \citet{buchanan2005no} discussed the problem of decay in certain organizational changes, i.e. ``initiative decay'' where the gains from change are lost when new practices are abandoned. 
According to a study by \citet{mendoza2010software}, the ease with which users can access ongoing training and their ability to perceive a technology's utility are the two main factors that encourage the productive use of a newly introduced technology. Later work by \citet{barki2007information} revealed that users tend to pick up adaptive behaviors in response to technological failures, which is beneficial to the organization in terms of efficiency by facilitating a better match between the system and the context in which it is being used.
However, given the research regarding text-based reporting systems, existing studies do not sufficiently address the question of how users engage with agents from safety reporting organizations when utilizing these text-based systems by analyzing empirical data, particularly during conversations.

\subsection{Emotional Supports with Conversational Systems}
The capability to provide emotional support is vital for various conversational agents, including those involved in clinical support \cite{denecke2020mental, wang2021cass, daher2020empathic}, customer service \cite{gelbrich2021emotional}, social interactions \cite{hu2018touch,meng2023mediated, laestadius2022too,xu2017new} and education \cite{ali2023supporting}. 
Within the domain of Human-Computer Interaction, the integration of emotional dialogue takes on pivotal significance in enriching the natural dynamics of human communication. 
The acknowledgment of the vital role of emotions in human interaction prompts the integration of automated emotional dialogue systems, intending to bridge this gap and cultivate more intuitive and meaningful interactions between users and computers. Recently, the emergence of LLMs has led researchers to scrutinize their effectiveness in emotion classification for diverse applications. Some researchers strategize methods to elicit emotional support from LLMs, employing strategies such as prompting \cite{li2023large, wang2023emotional} and data-augmented \cite{zheng2023augesc} fine-tuning.
However, the utilization of emotional chatbots in the safety reporting domain remains relatively unexplored, primarily due to constraints related to accessing large datasets and annotated emotional information. In the safety reporting domain, AI-mediated conversational interaction with users can assist dispatchers' handling of negative emotions such as despair and feelings of danger and mitigate emotional labor \cite{tracy1998emotion, wang2022desired}.

Despite the crucial role emotional chatbots could play in supporting users through challenging situations, their utilization in safety reporting remains relatively unexplored. 
To provide further understanding in this domain, we examined the text-based reporting process from a multi-stakeholder and longitudinal perspective that encompasses both dispatchers and users. 
Drawn from the findings of our analyses, we introduced and evaluated a fine-tuned LLM for the safety incident reporting domain.
The insights then consequently informed our design recommendations for future AI-enabled text-based risk reporting systems.

\yirenrevised{\section{Research Questions}}
\yirenrevised{
Given the above literature, we address the following research questions:
}

\begin{itemize}

    \item \textbf{RQ1}: \yirenrevised{\textit{When do users express negative emotions in text-based incident reports?}}

    \item \textbf{RQ2}:  \yirenrevised{\textit{How do dispatchers provide emotional support in response to different types of incidents?}}

    \item \textbf{RQ3}: \yirenrevised{\textit{How can LLMs be applied to enhance the perceived emotional support in the incident handling?}}

\end{itemize}

\yirenrevised{
In the remainder of this paper, we first present an overview of the methods employed for analyzing the system chat logs and  the development of the \textit{dispatcherLLM} model. We then detail the  results and findings yielded from the analysis and evaluation of the proposed model. Finally, we discuss the findings and the implications for the future design of text-based reporting systems. 
}

\section{Method}

To address the proposed research questions, we conducted a comprehensive system log analysis of a real text-based safety incident reporting system, called LiveSafe\footnote{https://www.vectorsolutions.com/solutions/vector-livesafe}, which has been used by more than 200 organizations, serving communities varying in location region and size. The chat log of each incident report consists of both users'  messages and dispatchers’ replies. The dataset allowed us to conduct the following analyses: 

\begin{itemize}
    \item To answer \textbf{RQ1}, what emotional states are presented in users' text reports of safety incidents, we analyzed the users' messages. Specifically, we applied emotion classification using the GoEmotions dataset \cite{demszky2020goemotions} and the calculation of Polarity Scores. We also executed statistical analyses to identify contextual factors that were associated with users' emotions in the reports. 
    \item To answer \textbf{RQ2}, whether dispatchers provide emotional support in their replies through the text-based reporting system, we conducted text analyses on the dispatchers' replies. We also employed statistical analyses to identify contextual factors that were associated with dispatchers’ delivery of emotional support.
    \item To answer \textbf{RQ3}, we fine-tuned the Llama-2 model \cite{touvron2023llama} to yield a \textit{dispatcherLLM} (Language Learning Model), which can be used to suggest replies by simulating human dispatchers' emotion support languages. We further completed an evaluation by comparing our proposed \textit{dispatcherLLM} with existing LLMs and showed the improved performance of delivery emotional support. \yirenrevised{We also presented results where participants rated their preferred responses and explained their choices. }
\end{itemize}

Below, We provide details of the dataset, the analysis procedure, and the method used for fine-tuning the \textit{dispatcherLLM} model.

\subsection{Dataset}
\begin{figure}[th!]
  \centering
     \begin{subfigure}[b]{0.2\textwidth}
         \centering
         \includegraphics[width=\textwidth]{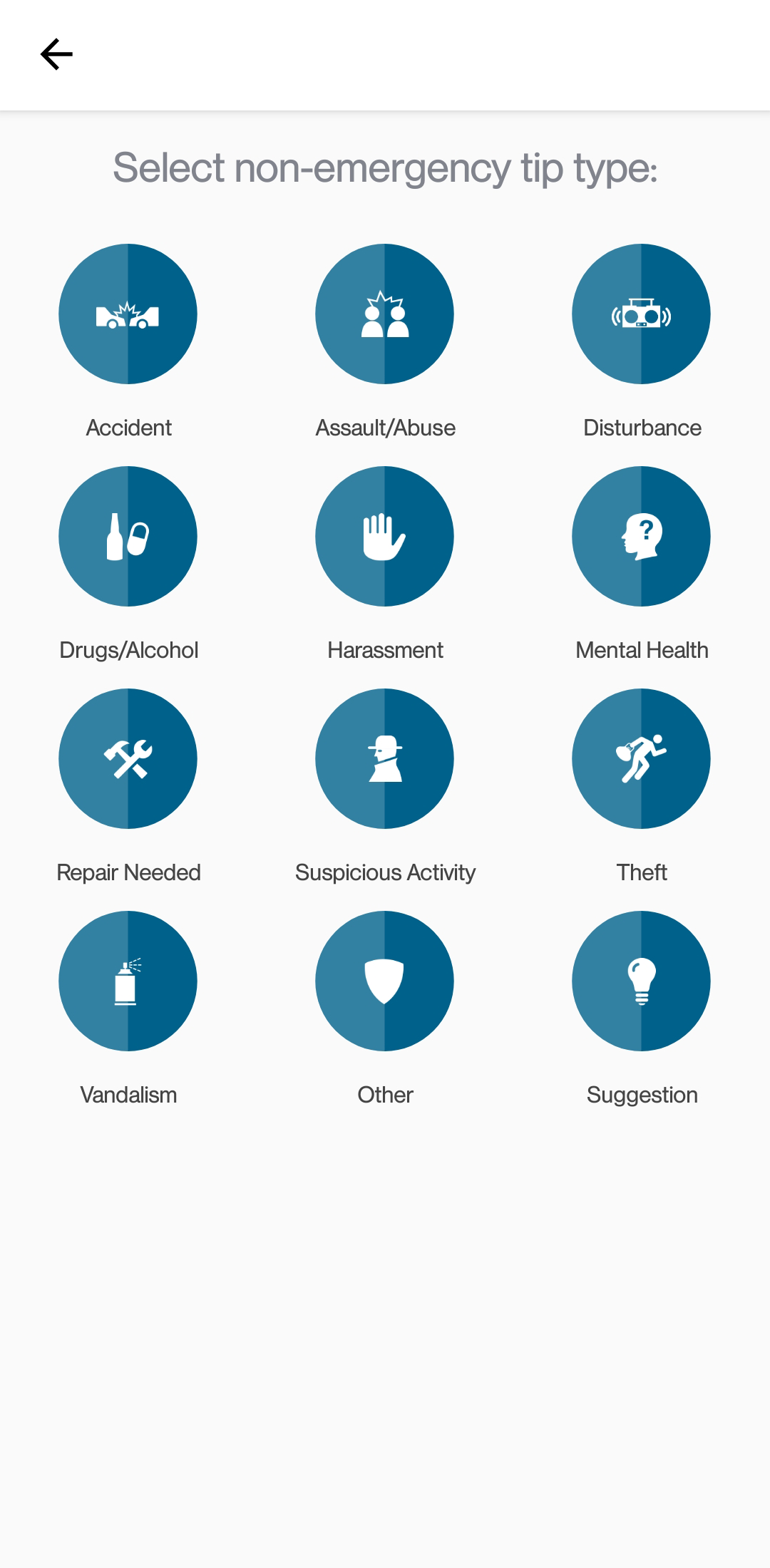}
         \caption{Tip Categories}
         \label{fig:livesafeInterface: tipCategory}
     \end{subfigure}
     \begin{subfigure}[b]{0.2\textwidth}
         \centering
         \includegraphics[width=\textwidth]{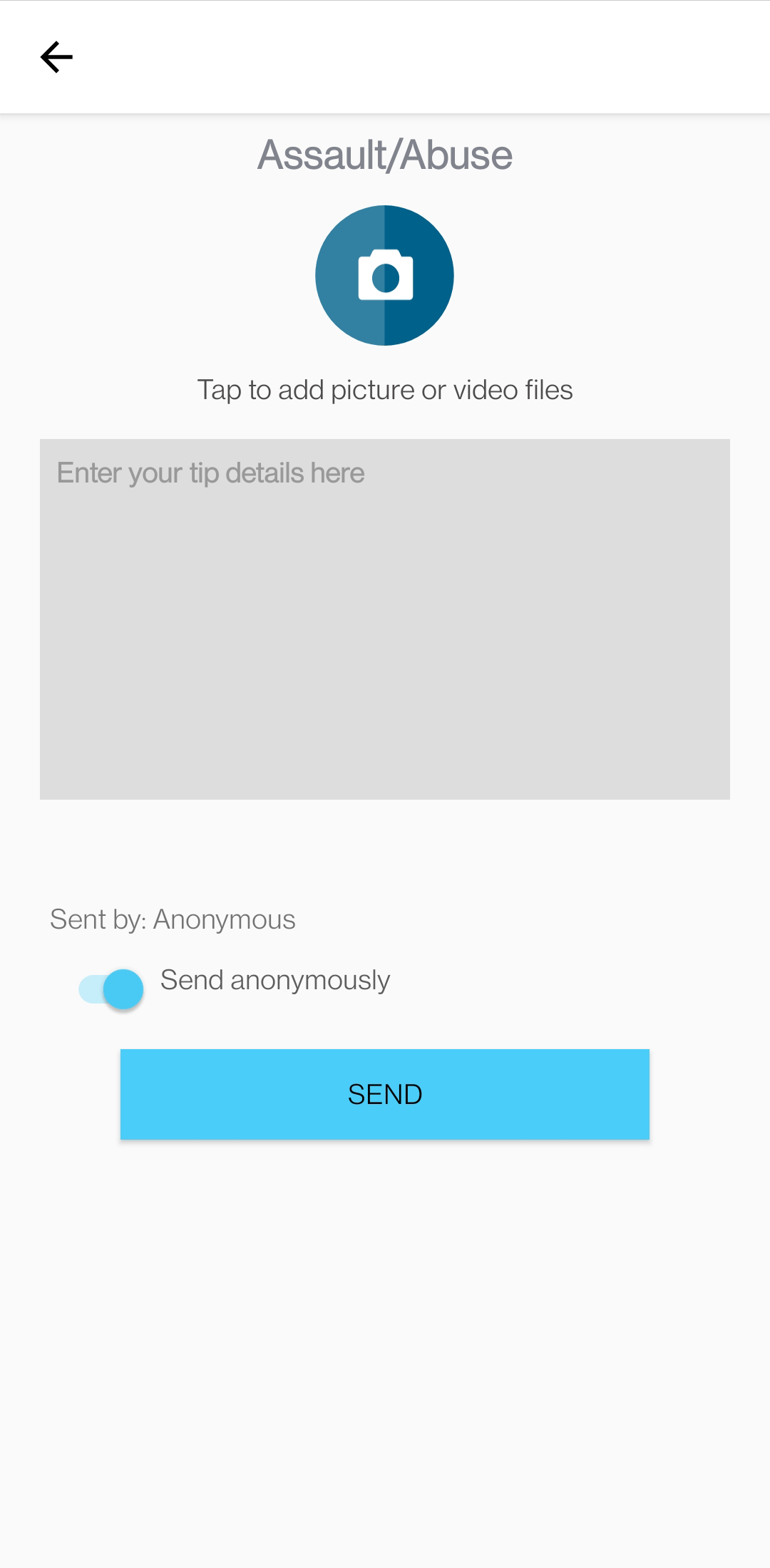}
         \caption{Submitting Tip}
         \label{fig:livesafeInterface: submitTip}
     \end{subfigure}
      \begin{subfigure}[b]{0.4\textwidth}
         \centering
         \includegraphics[width=\textwidth]{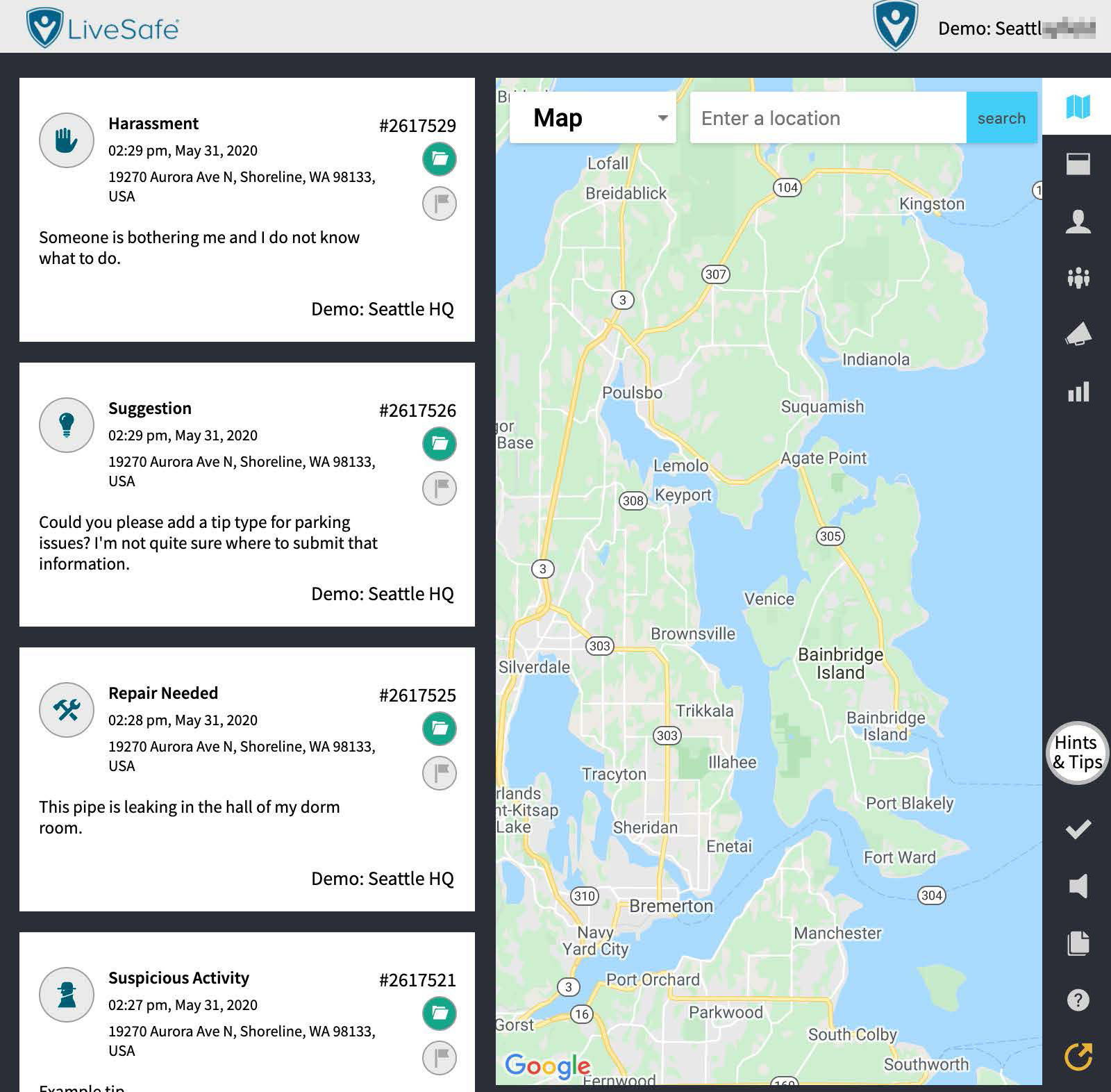}
         \caption{Dispatcher's Dashboard}
         \label{fig:livesafeInterface: dispatcherView}
     \end{subfigure}
  \caption{The user interface of the LiveSafe mobile app; 
  a) The user needs to select a tip category after choosing to submit a tip; 
  b) After the tip category is chosen, the user can submit text and image descriptions and connect to a human agent; 
  c) The dispatcher can connect and chat with the user using the Command Dashboard.
  }
  \label{fig:livesafeInterface}
\end{figure}

The chat log was collected through the LiveSafe app across the time period between 2018 and 2019, which allows users to report certain events to a human agent via a chat service.
LiveSafe is a risk management system that aims to facilitate communication between members of an organization and the safety teams responsible for managing risk \cite{ming2021examining}. This tool, which was first introduced in 2013, has been adopted by over 200 higher education institutions. Members of the community can submit information or "tips" through the LiveSafe mobile app or web portal, while safety organization dispatchers can respond to these tips through the LiveSafe Command Dashboard. Tips can take the form of text, photographs, video, or audio recordings and can be classified into different "tip types," which are chosen by local safety departments for each organization, as shown in Fig. \ref{fig:livesafeInterface}. After a user submits a tip, dispatchers from safety departments can engage in a conversation with the user through the Command Dashboard, as shown in Fig. \ref{fig:livesafeInterface: dispatcherView}).

\subsubsection{Data ethics}
Ethical considerations pertaining to data were taken into account before and throughout the entire research process due to the potential for incident reporting logs to contain private and sensitive information. 
Several steps have been taken to ensure that data ethics, such as anonymization, data minimization, and consent, were properly addressed.

First, the data was carefully anonymized by the organization before being accessible to the researchers for conducting the log analysis. 
The organizations and users were represented using anonymous IDs, and all mentions about geolocations, times and person/organization names in the chat log text were detected and replaced with tags (e.g., [LOCATION], [NAME]). 
As an additional measure of anonymization, a random selection of universities was further removed entirely from the dataset by LiveSafe. The dataset was fully anonymized.

Second, the researchers of this study did not actively collect the data, nor did they have access to any identifiable user information. The data was provided by the LiveSafe organization to researchers for analysis. An NDA (non-disclosure agreement) was established between the researchers and the LiveSafe organization. 
The study examined information from organizations and individual users who had consented to the use of anonymized and aggregated data for system improvement research.

Third, our research strictly adhered to the principle of data minimization, limiting our data collection to only what was essential for the study. 
All personal and sensitive information had been removed, and identifiable information, including user, geolocation, and organization information, had been removed or anonymized prior to researchers gaining access to the data.

\subsubsection{Data cleaning and pre-processing}
To reduce the noise within the chat log data, we first cleaned up the data by removing "Test" tips and irrelevant/ambiguous categories, which include incidents from categories including \textit{SafeRide; 911 / Call; Broadcast Message; Scavenger Hunt / Test; Operational Procedure Log; Broadcast Check-in Drill; Broadcast Check-in; Request Security Presence / Walk-through; Other; Misc}. 
We only kept tips after 2018 (between 2018-01-01 and 2019-12-31) and chats with more than 2 utterances (one conversation turn), since we want to investigate chats with at least one turn of information-collecting conversation.
Finally, for each tip we performed language detection using \textit{langdetect} \cite{nakatani2010langdetect} to identify messages in English. We identified 16.2\% (1,588 out of 9,827) of the tips to be non-English, and removed them to simplify further analysis. 



After performing cleaning, the resulting dialogues are categorized into different tip categories, which individuals selected before initiating a chat conversation. 
The dataset consists of 8,239 incidents, involving 18 different incident categories,  where the top 10 categories of incidents with highest tip counts are \textit{Noise Disturbance} (N=2,282), \textit{Suspicious Activity} (N=1,700), \textit{Emergency Message} (N=1,133), \textit{Drugs/Alcohol} (N=1,031), \textit{Facilities/Maintenance} (N=593), \textit{Harassment/Abuse} (N=455), \textit{Accident/Traffic/Parking} (N=348), \textit{Theft/Lost Item} (N=269), \textit{Mental Health} (N=171) and \textit{Vandalism/Damage} (N=90).
Sensitive information (names, places, times, etc) in the dataset was masked for privacy. 
It is worth noting the impacts this might have had on the annotation process and the model performance, as it is not always clear what the masked value was (e.g. "The incident must have occurred sometime between [TIME] and [TIME] which is when I got out of work.").

The dataset provides abundant conversation logs for further analysis. To distinguish it from the later annotated sample, we refer to this dataset as all data, containing a total of 8,239 conversations and 57,114 utterances.




\subsection{Detecting Emotions Involved in the Chats}
\label{sec: method_emotion}
In order to gain a deeper understanding of the emotional dynamics during conversations between users and dispatchers, we use an ML-based method to classify and quantify the emotions of the chat utterances. In this section, we discuss the methods used in this study to 1) classify the emotion of each chat utterance; 2) quantify and aggregate emotion negativity on a conversational level; and 3) identify dispatchers' delivery of emotional support.  

\subsubsection{Classifying emotions with RoBERTa and identifying emotional support} 
\label{sec:emotional_support_detection}
Following the approach of \cite{liu2023discovering}, we used a RoBERTa-based model to perform emotion classification to identify whether dispatchers provided emotional support \cite{vaux1987modes}. The RoBERTa model was finetuned using the GoEmotions dataset introduced by \citet{demszky2020goemotions}. The dataset contains 58K Reddit comments annotated with 28 types of fine-grained emotions (including \textit{neutral}). 
We refer to the classification of positive/neutral/negative emotions as described in the GoEmotions dataset.
The fine-tuned model was able to classify text into the 28 fine-grained emotion classes, reaching a satisfying overall accuracy of 93.5\% when evaluated on the GoEmotions Dataset.

To further identify the emotional support within dispatchers' utterances, we utilize the emotion classification model obtained above. We employed a similar method to identify emotional support as used in \citet{khanpour2018identifying}'s work within the domain of online health communities.
After applying the emotional classification model over all dispatchers' chat utterances (N = 33,718), we consider the utterances classified under the emotions, including   ``caring,'' ``love,'' ``sadness,'' ``remorse,'' and ``grief,'' as providing emotional support according to the definition of \citet{weber2004communcating}. 
For example, by applying the above method, we identified the following chats where the dispatcher showed emotional support to an incident labeled as \textit{Emergency Message}, where a dispatcher reassured the caller by texting, ``\textit{I'm with you and officer [PERSON] and officer [PERSON] will be right there I promise.}'', which was classified as caring by the model.
In a different scenario, the dispatcher texted “\textit{Ok. I am sorry to hear that happened to you,}” when the user reported a  \textit{Theft / Lost Item} incident, which was identified by the model as sadness.

\subsubsection{Measuring conversational emotion polarity}
\label{negativity_metric}  
To further measure the emotional polarity of users throughout conversations, we quantify and aggregate the negativity of users' utterance emotions by conversation specifically for the context of safety reporting. 
We identified a number of heuristics to better capture the traits of the safety reporting domain after qualitatively analyzing the chat logs.  Among these heuristics are: 1) The majority of users have bad emotions when they first start a chat, and dispatchers usually fix or settle them gradually throughout conversations. 2) If a user still has negative emotions at the end of the session, that is more concerning.

Drawing from existing literature \cite{zhang2009sentiment, ghandeharioun2019approximating}, we use a measurement metric for quantifying the negativity of users' utterances throughout a single reporting conversation by calculating the relative frequency of users' utterances with negative emotions. We define negative emotions as in the work of  \citet{demszky2020goemotions}, which consists of fine-grained emotions including \textit{anger}, \textit{annoyance}, \textit{disappointment}, \textit{disapproval}, \textit{disgust}, \textit{embarrassment}, \textit{fear}, \textit{grief}, \textit{nervousness}, \textit{remorse} and \textit{sadness}. 
We modified the emotion polarity score metric with an increasing weight toward the end of each conversation based on our heuristics, which can be denoted as:
\begin{equation}
    S_n = \frac{\sum_i^N{s_ie^i}}{\sum_i^N{e^i}}
\end{equation}
where $n$ denotes the $n^{th}$ conversation with a total count $N$ of user utterances. $s_i$ refers to the emotional polarity of the $i^th$ user utterance, i.e. -1 for negative emotions and 0 for neutral and positive emotions. The resulting polarity score $S_n$ represents the user's level of negative emotion during the $n^{th}$ conversation. 
For example, with user utterances with emotions [fear, confusion, neutral, curiosity, gratitude], $s$ will be [-1, -1, 0, 0, 0].
The resulting polarity score is denoted as $S \in [-1, 0]$, indicating how negative the user's overall emotion is throughout the reporting conversation.

\subsection{Measuring Information Collection through Event Argument Extraction}
In order to further analyze how dispatchers collect incident-related information from users during conversations, we introduced a BERT-based model to automatically extract event argument information taking the conversation utterance text as input. 
In the following section, we first introduce how an event ontology was introduced for safety reporting purposes. Then, we describe the model used for automatic event argument extraction, which was fine-tuned and evaluated using our proposed event ontology. 

\subsubsection{Event ontology} 
\label{Annotation_method}
To further understand how dispatchers collect incident-related information from users during conversations, we refined the event argument schema based on the Automatic Content Extraction (ACE) dataset \cite{walker2006ace} and adapted it to the domain of safety incident reporting. The ACE dataset is an event extraction task dataset widely used by prior studies \cite{lin2020joint, xia2019multi, li2021document}. 
The dataset in question comprises annotations pertaining to event and entity information obtained from news articles. 
While the dataset does partially encompass events that are relevant to the domain of safety, such as \textit{Attack} with arguments including \textit{Attacker}, \textit{Target}, \textit{Instrument}, \textit{Time}, and \textit{Place}, the original ACE dataset schema falls short in its ability to directly annotate incidents related to safety due to its coarse-grained nature. To address this issue, we have modified the ACE event schema and subsequently employed it in the analysis of LiveSafe chat logs. 
Based on the initial ACE event argument schema, an exploratory annotation process was undertaken by the researchers. During the annotation process, we identified a set of event-related arguments, including \textit{Attacker}, \textit{Target}, \textit{Location}, \textit{Weapon}, \textit{Start Time}, \textit{End Time}, and \textit{Target Object}. These arguments were linked to the event handle, constituting a vital component of the event information. The complete ontology and definition can be found in Appendix \ref{appendix}.





\subsubsection{Model}
\label{Event_argument_extraction_method}
To capture the event information from the conversations between users and dispatchers, we established an ML-based model to perform event argument extraction. The model is based on the safety reporting event ontology introduced in prior work by \citet{liu2023discovering}.
Our event argument extraction model aims to extract certain information about a safety incident given the dialogue history. 
In order to achieve this, we have identified and selected key fields that are related to the categories of interest, as in the ontology obtained in \ref{Annotation_method}, and can be extracted as spans of text from the dialogue history. Our model design adheres to the zero-shot setting of STARC \cite{gao-machine-2020}, which is a state-of-the-art method that can be applied to new domains without additional fine-tuning. 

The system design involves the choice of questions for each field, the QA model selection, and the decoding of QA outputs into event argument extraction outputs. Several phrasings of natural language questions were compared for each slot. For example, for the "object stolen" slot on theft, the questions "What was stolen?" and "What object was stolen?" were compared, and the second phrasing scored higher. 
For every new utterance in the dialogue, we predict a new dialogue state consisting of a list of predicted values for each pre-defined slot. The dialogue history until the utterance is passed to the QA model, along with one question for each slot. Valid answers (controlled by the hyper-parameters of a minimum score and maximum length) are appended to an ongoing list of answers for each slot. 
 

\subsection{\textit{DispatcherLLM}: A LLM for Generating Scenario-Based Dispatcher Responses}
\label{sec:dispatcherLLM}
\yirenrevised{
To explore the feasibility of improving both informational and emotional support of incident report handling using AI, we fine-tuned an LLM on safety reporting chat logs from \textit{LiveSafe}. The model's generation output is evaluated using both automatic metrics and human evaluation with community members. 
}
\subsubsection{Instruct-tuning with chat log}
\yirenrevised{
A Llama-2 model \cite{touvron2023llama} was fine-tuned and evaluated using the same set of chat logs analyzed in this study. We used the 7B instruction-tuned version of the Llama-2 model released by Meta. We chose the instruct-tuned Llama-2 7B model because of its open-sourced nature and ability to be replicated and validated in various application scenarios. Meanwhile, the instruct-tuned Llama-2 model has been aligned to handle user instructions and deliver emotional support for general applications \cite{huang2023emotionally}. In this study, we further fine-tuned the model to adapt to the task space of incident report handling while utilizing the model's learned ability to provide emotional support.
}
The model is fine-tuned on a training partition of 31,359 dispatcher utterances. A validation set consisting of 10,453 dispatcher utterances was used to assess the generalization of the trained model. We refer to the resulting model as \textit{dispatcherLLM}.
Since the instruct-tuned model should be capable of accommodating various incident scenarios, we generate a summary of incident scenarios using GPT-3.5 for each chat log. The generated summaries were then incorporated as scenario descriptions during the fine-tuning process. 
During supervised fine-tuning, we fine-tuned the model to generate each dispatcher response given the incident scenario summary and dialogue history. Specifically, dialogue history and incident summary are pre-filled as pre-formatted prompts for Llama-2 and then aligned with the dispatcher's utterance by each turn.
In the inference phase, our system retrieves the most similar document from the training set, based on both the summary and dialogue history, in order to enhance the \textit{dispatcherLLM}'s responses. An example of the constructed prompt and dispatcher responses generated by the model can be found at \ref{appdx:chat_example}.

\subsubsection{\yirenrevised{Human evaluation}}
\yirenrevised{
To further evaluate the delivery of effective emotional support, we conducted a survey study with community members from a higher education institution. We recruited 17 undergraduate students from a U.S.-based university community and designed a survey to collect feedback by asking them to compare responses generated by the instruct-tuned model with both responses generated by GPT-3.5 and human dispatchers' responses. The participants are instructed to compare the outputs from GPT-3.5 and \textit{dispatcherLLM} with human dispatchers' responses and choose the better response based on different scenarios randomly sampled from the test set. 
We also collected users' qualitative feedback and ratings on the emotional support provided by the two models' generated responses. The findings of the survey study are presented in \ref{sec:emotional_support_findings}.
The survey details and included samples can be found in the appendix (\ref{apdx:survey_cases} and \ref{apdx:survey_questions}). This study aimed to gather human feedback on the generated emotional support to inform future research on model enhancement and practical applications. It is not intended to be exhaustive or conclusive.
}

\section{Findings}

\subsection{Users' Emotions Vary by Incident Type and Reporting Stage (RQ1)}


We first examine the emotional dynamics within the conversation between users and dispatchers. In order to conduct a more comprehensive analysis, we base our analysis on the results obtained from the emotion detection model described in \ref{sec: method_emotion}. 
Using the measurement introduced in \ref{negativity_metric}, we calculated the emotional polarity score for all incident reporting conversations. The resulting polarity scores ranged between -0.75 and 0.
The results showed that while most tips (N=6,623, 82.7\%) contained no negative emotions from users' utterances, tips with extreme polarity scores less than or equal to -0.5 (N=262, 3.3\%) were also found. 
To further understand what led to users' negative emotions, we conducted analysis to identify the different contextual factors, including tip category, anonymity, years in use and tip volumes, associated with users' emotional statuses.

\subsubsection{
Users expressed more negative emotions for certain incident types
}
For most tip categories, the first messages from users were often neutral in emotion.
This suggests that users may initially approach the dispatcher with a calm and collected demeanor in the majority of incident types reported.
The results also show that users tend to express more gratitude for incidents that are less urgent in nature, such as facilities/maintenance, noise disturbance, theft/lost item, and vandalism/damage. This may be because these types of incidents are typically less severe and do not require immediate action, allowing the users to feel more grateful for the assistance provided by the dispatcher.
In contrast, for incidents related to mental health and injury/medical, users tend to express more caring emotions.
This may be because these types of incidents are often more serious and require a more compassionate response from the dispatcher. The higher level of caring emotions in these situations may reflect the users' concern for their own well-being or the well-being of others involved in the incident. 



\begin{figure}[!h]
  \centering
  \includegraphics[width=0.9\linewidth]{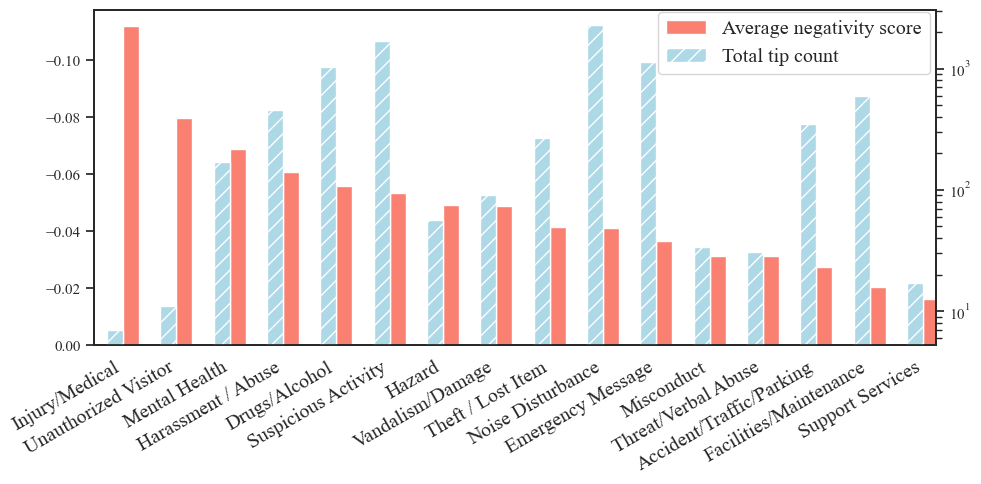}
  \caption{Average negativity score and total tip count by tip category. 
  ANOVA revealed a significant difference ($F(17, 8221) = [4.65], p < 0.001^{***}$) in user's emotional polarity across incident categories during reporting.
  }
  \label{fig:emotion_score_dist}
\end{figure}

As shown in Figure \ref{fig:emotion_score_dist}, it is also found that the average polarity scores differ across different incident categories. 
A one-way ANOVA indicated a significant disparity across incident categories ($F(17, 8221) = [4.65], p < 0.001^{***}$).
Incidents that are more urgent in nature or related to personal safety tend to incur more exhibited negative emotions from users ($t(722) = [-2.13], p = 0.034^{*}$), e.g. \textit{Harassment / Abuse} ($M=-0.06, SD = 0.13$) vs. \textit{Theft/Lost Item} ($M = -0.04, SD = 0.11$). Meanwhile, \textit{Mental Health} ($M = -0.07, SD = 0.14$) related incidents have also been found to have more negative emotions in users' utterances during reporting. 

\begin{table}[ht!]
\resizebox{\linewidth}{!}{
\begin{tabular}{lllllll}
\hline
                                       & \multicolumn{2}{l}{Model 1}        & \multicolumn{2}{l}{Model 2}          & \multicolumn{2}{l}{Model 3}          \\
                                       & \textbf{Coeff.}  & \textbf{(S.E.)} & \textbf{Coeff.}    & \textbf{(S.E.)} & \textbf{Coeff.}    & \textbf{(S.E.)} \\ \hline
\textbf{Event category}                &                  &                 &                    &                 &                    &                 \\
\hspace{3mm} Suspicious activity       & REF              &                 &                    &                 &                    &                 \\
\hspace{3mm} Accident/Traffic/Parking  & -0.17            & (0.35)          & -0.15              & (0.41)          & -0.15              & (0.40)          \\
\textbf{\hspace{3mm} Contact Mall/Corporate/Property Security} &
  \textbf{-38.51} &
  \textbf{(0.24)} &
  \textbf{-17.97**} &
  \textbf{(0.48)} &
  \textbf{-13.93**} &
  \textbf{(0.48)} \\
\hspace{3mm} Drugs/Alcohol             & 0.02             & (0.23)          & -0.01              & (0.22)          & -0.02              & (0.23)          \\
\textbf{\hspace{3mm} Emergency Message} &
  \textbf{0.81***} &
  \textbf{(0.18)} &
  \textbf{0.81***} &
  \textbf{(0.25)} &
  \textbf{0.79***} &
  \textbf{(0.25)} \\
\hspace{3mm} Facilities/Maintenance    & 0.30             & (0.24)          & 0.31               & (0.22)          & 0.29               & (0.22)          \\
\hspace{3mm} Harassment/Abuse          & 0.30             & (0.27)          & 0.28               & (0.24)          & 0.27               & (0.24)          \\
\hspace{3mm} Hazard                    & 0.08             & (0.73)          & 0.08               & (0.81)          & 0.01               & (0.81)          \\
\hspace{3mm} Injury/Medical            & 1.51             & (1.09)          & 1.55               & (0.97)          & 1.46               & (0.97)          \\
\textbf{\hspace{3mm} Mental Health}    & \textbf{1.97***} & \textbf{(0.24)} & \textbf{1.95***}   & \textbf{(0.26)} & \textbf{1.96***}   & \textbf{(0.26)} \\
\textbf{\hspace{3mm} Misconduct}       & \textbf{-16.09}  & \textbf{(0.67)} & \textbf{-23.88*} & \textbf{(0.50)} & \textbf{-23.85*} & \textbf{(0.50)} \\
\hspace{3mm} Noise Disturbance         & -0.28            & (0.20)          & -0.33              & (0.24)          & -0.31              & (0.24)          \\
\textbf{\hspace{3mm} Support Services} & \textbf{1.74**} & \textbf{(0.66)} & \textbf{1.76.}     & \textbf{(0.95)} & \textbf{1.71.}     & \textbf{(0.94)} \\
\textbf{\hspace{3mm} Suspicious/Unattended Package} &
  \textbf{-14.80} &
  \textbf{(0.86)} &
  \textbf{-15.38**} &
  \textbf{(0.82)} &
  \textbf{-18.64**} &
  \textbf{(0.82)} \\
\textbf{\hspace{3mm} Theft/Lost Item} &
  \textbf{0.88**} &
  \textbf{(0.27)} &
  \textbf{0.85**} &
  \textbf{(0.27)} &
  \textbf{0.87**} &
  \textbf{(0.27)} \\
\textbf{\hspace{3mm} Threat/Verbal Abuse} &
  \textbf{-1.90} &
  \textbf{(2.49)} &
  \textbf{-59.56*} &
  \textbf{(0.33)} &
  \textbf{-31.56*} &
  \textbf{(0.33)} \\
\hspace{3mm} Unauthorized Visitor      & 1.16             & (1.06)          & 1.25               & (0.93)          & 1.19               & (0.93)          \\
\hspace{3mm} Vandalism/Damage          & 0.01             & (0.60)          & -0.02              & (0.74)          & -0.05              & (0.75)          \\
Anonymous                     & -0.39            & (0.31)          & -0.40              & (0.33)          & -0.21              & (0.14)          \\
Time of day                            &                  &                 &                    &                 &                    &                 \\
\hspace{3mm} 4 a.m. - 8 a.m.           & -0.41.          & (0.19)          & -0.41.            & (0.16)          & -0.40              & (0.33)          \\
\textbf{\hspace{3mm} 8 a.m. - 12 p.m.} & \textbf{0.14}    & \textbf{(0.18)} & \textbf{0.13}      & \textbf{(0.19)} & \textbf{-0.41.}   & \textbf{(0.16)} \\
\hspace{3mm} 12 p.m. - 4 p.m.          & -0.03            & (0.20)          & -0.03              & (0.18)          & 0.14               & (0.19)          \\
\hspace{3mm} 8 p.m. - 12 a.m.          & -0.12            & (0.17)          & -0.12              & (0.15)          & -0.04              & (0.18)          \\
\hspace{3mm} 12 a.m. - 4 a.m.          & 0.00             & (0.00)          & 0.00               & (0.00)          & -0.13              & (0.16)          \\
chat length                            & -0.19            & (0.12)          & -0.20              & (0.14)          & 0.00               & (0.00)          \\
Org. years in use                 &                  &                 & 0.07               & (0.05)          & 0.08               & (0.05)          \\
Org. tips received per year            &                  &                 &                    &                 & 0.00*              & (0.00)          \\ \hline
{
\scriptsize . p < 0.05; * p < 0.01; ** p < 0.005; *** p < 0.001 (two-tail); S.E. clustered by organization}
\end{tabular}
}
\caption{
The results of a linear regression present the factors that are associated with users' emotional negativity (N = 8,239). The higher the coefficient, the more likely users were to express negative emotions (e.g., for mental health and emergency incidents);
Models 1, 2, and 3 gradually introduced new variables (anonymity, years in use, and tip volume) to demonstrate robustness. 
}
\label{Tab:lr_Negativity}
\vspace{-10pt}
\end{table}

To further examine the factors associated with users' negative emotions, we conducted a linear regression analysis over whether users' first message was emotionally negative. 
As shown in Table \ref{Tab:lr_Negativity}, we found that users' negative emotions are associated with certain categories of events, such as \textit{Emergency Message} and \textit{Mental Health}. 
The results of the regression revealed no association between users' negative emotions and organizations' years in use of the LiveSafe system, and the average number of tips received per year by the organization. 
The findings have shown that users' negative emotions only varied by the factors related to the nature of incidents and the design choices of the reporting systems, but not by characteristics of the organizations, which is reflected through the years in use since implementation (experience) and average tips per year (usage and scale of the system). 



\subsubsection{Users express more positive emotions with dispatchers involved during conversations}

\begin{figure}[!t]
  \centering
  \includegraphics[width=0.8\linewidth]{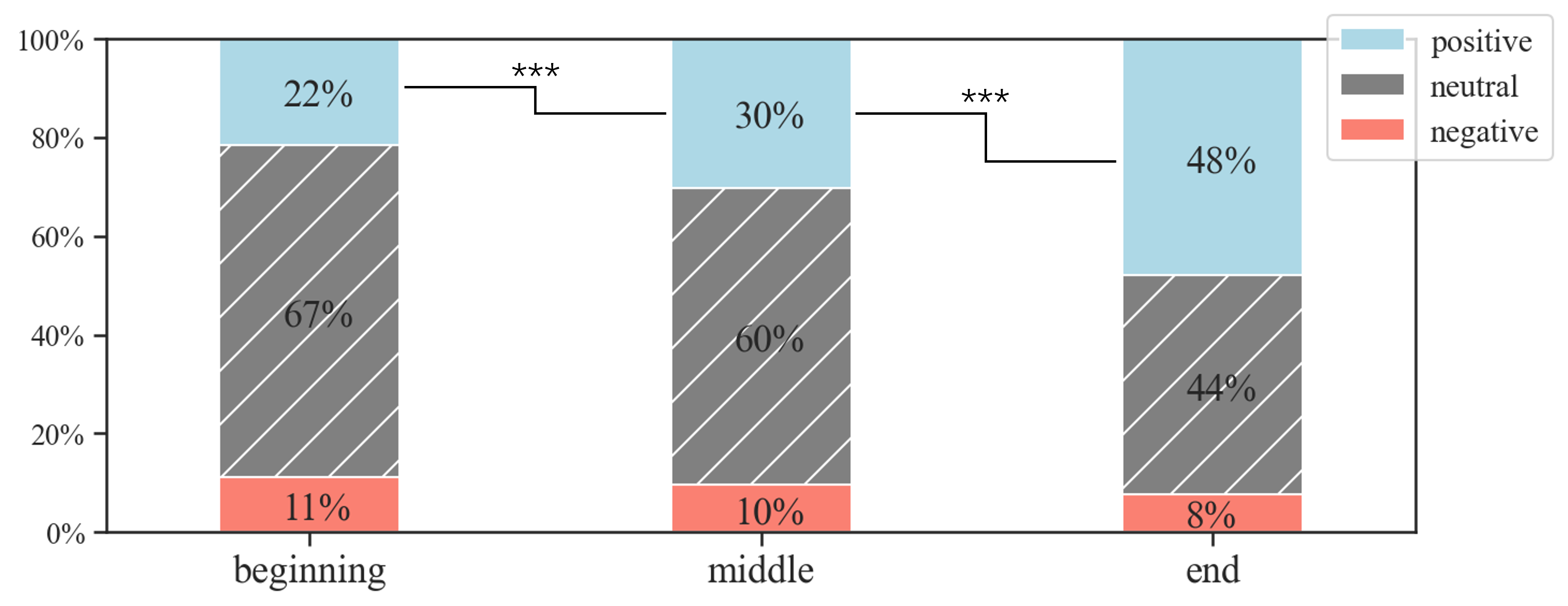}
  \caption{Change in the sentiment ratio of users' utterances throughout conversations: \textit{\textbf{beginning}} refers to the first 1/3 of each conversation, \textit{\textbf{middle}} refers to the 2/3 of each conversation, and \textit{\textbf{end}} refers to the last 1/3 of each conversation. 
  The t-tests indicate a significant increase in users' positive sentiment as their conversations with dispatchers progressed.
  } 
  \label{fig:emotion_ratio}
\end{figure}

During the reporting conversations, users first provide an overall description of the incidents. Dispatchers then join the conversations to further collect more information and provide assistance to address users' concerns. 
To further understand users' emotion changes during this process, we conducted a sentiment analysis based on the emotion classification results described in \ref{sec: method_emotion}. 
Each conversation is divided into three stages --- the initiation stage (first 1/3 of utterances), the information-gathering stage (middle 1/3 of utterances), and the elaboration stage (last 1/3 of utterances).
As shown in the results of the emotion analysis in Figure \ref{fig:emotion_ratio}, we discovered an increase in users' positive sentiment as conversations proceeded. 
A Chi-Square Goodness of Fit Test showed that the sentiment distributions exhibited significant differences among the three stages of conversations ($ X^2 (4, 8239) = [1160.34], p < 0.001^{***} $). 
This is reflected in a consistent increase in the ratio of utterances with positive sentiment ($t = [-31.369], p < 0.001^{***}$), and a decrease in ratios of utterances with both neutral and negative sentiment. 
The decrease in users' neutral utterances implied the process of users providing incident-related information during the earlier stage of conversations, which later led to a higher proportion of positive utterances as dispatchers provided assistance after information was collected. 
Although a decrease in negative sentiment is observed, the magnitude of the change is not as obvious, which might imply a higher difficulty for dispatchers to address users' negativity in sentiment or a lack of focus in providing emotional assistance during incident report handling. 

\textbf{Summary  (\textbf{RQ1})}: 
In terms of users' emotions conveyed in their incident report messages, our findings showed that when reporting specific types of incidents, such as \textit{Mental Health} and \textit{Emergency Message}, users were more likely to start with negative emotions, as indicated in Table~\ref{Tab:lr_Negativity}. 
However, their emotions tended to become less negative towards the end of their interactions with the dispatchers, as shown in Fig.~\ref{fig:emotion_ratio}. 

\subsection{Factors Associated with the Delivery of Dispatchers' 
 Emotional Support (RQ2)}
\label{sec:emo_support_categories}

Whether dispatchers decided to provide emotional support to users during reporting conversations depends on many factors. Since this is often not included as a standard procedure in safety departments' playbooks, it is vital to understand what factors were associated with dispatchers' decision making in providing emotional support in practice. 
To examine this, we conducted a logistic regression analysis over the relationship between whether a dispatcher provided emotional support and a range of potential factors involved in the decision-making process. 

\begin{table}[th!]
\resizebox{\linewidth}{!}{
\begin{tabular}{lllllll}
\hline
                                                      & \multicolumn{2}{l}{\textbf{Model 1}} & \multicolumn{2}{l}{\textbf{Model 2}} & \multicolumn{2}{l}{\textbf{Model 3}} \\
\textbf{}                                             & \textbf{Coeff.}    & \textbf{(S.E.)} & \textbf{Coeff.}    & \textbf{(S.E.)} & \textbf{Coeff.}    & \textbf{(S.E.)} \\ \hline
\textbf{Event category}                               &                    &                 &                    &                 &                    &                 \\
\hspace{3mm} Suspicious activity                      & REF                &                 &                    &                 &                    &                 \\
\hspace{3mm} Accident/Traffic/Parking                 & -0.32              & (0.20)          & -0.30              & (0.20)          & -0.27              & (0.19)          \\
\hspace{3mm} Contact Mall/Corporate/Property Security & 0.22               & (1.10)          & 0.13               & (1.13)          & 0.18               & (1.13)          \\
\textbf{\hspace{3mm} Drugs/Alcohol}                   & \textbf{-0.52.}  & \textbf{(0.16)} & \textbf{-0.47.}  & \textbf{(0.16)} & \textbf{-0.47.}  & \textbf{(0.16)} \\
\textbf{\hspace{3mm} Emergency Message}               & \textbf{0.64***}   & \textbf{(0.12)} & \textbf{0.63***}   & \textbf{(0.12)} & \textbf{0.64***}   & \textbf{(0.12)} \\
\hspace{3mm} Facilities/Maintenance                   & 0.08               & (0.21)          & 0.11               & (0.21)          & 0.14               & (0.21)          \\
\textbf{\hspace{3mm} Harassment/Abuse}                & \textbf{0.64***}   & \textbf{(0.18)} & \textbf{0.68***}   & \textbf{(0.19)} & \textbf{0.67***}   & \textbf{(0.18)} \\
\hspace{3mm} Hazard                                   & -0.69              & (0.42)          & -0.58              & (0.44)          & -0.58              & (0.45)          \\
\hspace{3mm} Injury/Medical                           & 0.50               & (1.17)          & 0.41               & (1.18)          & 0.37               & (1.14)          \\
\textbf{\hspace{3mm} Mental Health}                   & \textbf{0.54.}   & \textbf{(0.21)} & \textbf{0.58.}   & \textbf{(0.20)} & \textbf{0.57.}   & \textbf{(0.21)} \\
\textbf{\hspace{3mm} Misconduct}                      & \textbf{0.98*}   & \textbf{(0.33)} & \textbf{0.88.}    & \textbf{(0.35)} & \textbf{0.91.}    & \textbf{(0.36)} \\
\textbf{\hspace{3mm} Noise Disturbance}               & \textbf{-0.42.}   & \textbf{(0.19)} & \textbf{-0.45*}  & \textbf{(0.16)} & \textbf{-0.43*}  & \textbf{(0.16)} \\
\hspace{3mm} Support Services                         & 0.06               & (0.60)          & 0.10               & (0.65)          & 0.14               & (0.64)          \\
\textbf{\hspace{3mm} Suspicious/Unattended Package} & \textbf{-16.90*} & \textbf{(0.76)} & \textbf{-19.74*} & \textbf{(0.77)} & \textbf{-18.00*} & \textbf{(0.77)} \\
\textbf{\hspace{3mm} Theft/Lost Item}                 & \textbf{0.44.}    & \textbf{(0.19)} & \textbf{0.47*}   & \textbf{(0.18)} & \textbf{0.49*}   & \textbf{(0.18)} \\
\textbf{\hspace{3mm} Threat/Verbal Abuse}             & \textbf{1.17**}   & \textbf{(0.39)} & \textbf{1.10**}   & \textbf{(0.42)} & \textbf{1.13**}   & \textbf{(0.41)} \\
\hspace{3mm} Unauthorized Visitor                     & 0.11               & (0.93)          & -0.05              & (0.95)          & -0.11              & (1.01)          \\
\textbf{\hspace{3mm} Vandalism/Damage}                & \textbf{-1.15.}   & \textbf{(0.51)} & \textbf{-1.04.}   & \textbf{(0.52)} & \textbf{-1.03.}   & \textbf{(0.52)} \\
\textbf{Anonymous}                                    & \textbf{-0.38***}  & \textbf{(0.09)} & \textbf{-0.38***}  & \textbf{(0.09)} & \textbf{-0.40***}  & \textbf{(0.09)} \\
\textbf{Time of day}                                  &                    &                 &                    &                 &                    &                 \\
\hspace{3mm} 4 a.m. - 8 a.m.                          & REF                &                 &                    &                 &                    &                 \\
\textbf{\hspace{3mm} 8 a.m. - 12 p.m.}                & \textbf{-0.65***}  & \textbf{(0.25)} & \textbf{-0.66***}  & \textbf{(0.23)} & \textbf{-0.66***}  & \textbf{(0.23)} \\
\textbf{\hspace{3mm} 12 p.m. - 4 p.m.}                & \textbf{-0.51***}  & \textbf{(0.20)} & \textbf{-0.53***}  & \textbf{(0.19)} & \textbf{-0.53***}  & \textbf{(0.19)} \\
\hspace{3mm} 8 p.m. - 12 a.m.                & -0.34    & (0.20) & -0.35    & (0.19) & -0.35    & (0.19) \\
\hspace{3mm} 12 a.m. - 4 a.m.                         & -0.32              & (0.23)          & -0.32              & (0.22)          & -0.32              & (0.22)          \\
\hspace{3mm} 4 a.m. - 8 a.m.                          & -0.12              & (0.20)          & -0.12              & (0.19)          & -0.14              & (0.19)          \\
\textbf{Org. years in use
}                & \textbf{}          & \textbf{}       & \textbf{-0.13***}  & \textbf{(0.03)} & \textbf{-0.13***}  & \textbf{(0.03)} \\
Org. Tips received per year                           &                    &                 & 0.00              & (0.00)          & 0.00               & (0.00)          \\
\textbf{User's emotional polarity}                      & \textbf{}          & \textbf{}       & \textbf{}          & \textbf{}       & \textbf{-1.13***}  & \textbf{(0.33)} \\ \hline
{
\scriptsize . p < 0.05; * p < 0.01; ** p < 0.005; *** p < 0.001 (two-tail); S.E. clustered by organization
}
\end{tabular}
}
\caption{
Results of a logistic regression analysis present the factors that are associated with emotional support delivered for incident reports (N = 8,239). For example, dispatchers were more likely to provide emotional support when users reported Harassment/Abuse incidents and less likely when users reported Suspicious/Unattended Package incidents. Meanwhile, dispatchers were less likely to provide emotional support during normal working hours between 8 am to 4 pm. 
Models 1, 2, and 3 gradually introduced new variables (years in use, tip volume, and users' emotion polarity score) to demonstrate the robustness.
}
\label{Tab: emotional_support_Logistic}
\vspace{-15pt}
\end{table}

\subsubsection{Tip-related factors} As shown in Table~\ref{Tab: emotional_support_Logistic}, the delivery of dispatchers' emotional support varied by several tip-related factors, e.g., tip categories, anonymity of the tips, and emotional polarity of the users' first messages. 
An inconsistency of emotional support across different tip categories was revealed by the results of the logistic regression. For example, users reporting incidents under the categories of \textit{Harassment/Abuse} and \textit{Mental Health} were found to be more likely to receive dispatchers' emotional support compared to \textit{Drugs/Alcohol} and \textit{Noise Disturbance}. 
Meanwhile, tips where users reported anonymously were found less likely to be provided with emotional support. 
We also found that users with a stronger negative emotion display in their messages were more likely to receive emotional support from dispatchers. 

\subsubsection{Service-related factors}
In order to gain deeper insight into the distribution of reported tips across various time periods, a Chi-Square Goodness of Fit Test was conducted to test the difference in the distribution of tips throughout different segments of the day. The results of the test indicated that there is a significant difference in the distribution of user-reported incident types among the different time periods of the day ($ X^2 (30, 8239) = [1245.21], p < 0.001^{***} $).
More specifically, higher volumes of incidents were being reported during periods of \textit{Afternoon}, \textit{Evening}, \textit{Night}, and \textit{Late Night}. The increase in volume is especially observable during the \textit{Night} with a drastic increase in incidents related to \textit{Noise Disturbance} and \textit{Drugs/Alcohol}. An increased amount of \textit{Harassment/Abuse} incidents was also observed during \textit{Afternoon}, \textit{Evening}, and \textit{Night}, while significantly smaller than the amount of other incidents such as \textit{Noise Disturbance} and \textit{Drugs/Alcohol}.



The results also showed that two service-related factors were associated with the delivery of emotional support, as shown in Table \ref{Tab: emotional_support_Logistic}. 
The results from Model 1 of Table \ref{Tab: emotional_support_Logistic} also showed additional inconsistency in emotional support based on when the report happened with respect to the time of day. Generally, we found that dispatchers tended to provide less emotional support during normal working hours during the day (i.e., 8 a.m. - 12 p.m. and 12 p.m. - 4 p.m.). 
Since there might be a difference in tip volumes of different tip categories reported during each time period of the day, it might affect dispatchers' behavior and strategies when handling users' incident reports. 
Due to the different volumes of incidents during each time period, dispatchers on each shift might experience burnout at varied levels. 
The findings suggested that shifts with higher incident report volumes could lead to greater dispatchers' burnout (e.g., \textit{Afternoon}, \textit{Evening} and \textit{Night}), thus decreasing the willingness for dispatchers to provide emotional support. 
The findings also revealed that dispatchers' differences in the delivery of emotional support might be affected by the volume of different reported tip categories at their corresponding shifts of the day. For example, dispatchers during \textit{Night} shifts might tend to deliver less emotional support since increased reports of \textit{Drugs/Alcohol} incidents. As shown by the logistic regression analysis results in Table \ref{Tab: emotional_support_Logistic}, this effect impacts not only the emotional support delivery of the increased tip category, but also other incident reports under the same time period/shift.

\subsubsection{Organizational Factors.}
\label{sec:emotional_decrease_finding} 
Also, we found that the likelihood of dispatchers providing emotional support to users was negatively associated with the amount of time that passed since organizations implemented their incident reporting systems. This might suggest that over time, dispatchers become more focused on resolving incidents and less likely to provide emotional support, which further indicates a possible decay in dispatchers' awareness of the delivery of emotional support when handling users' incident reports.
This effect was also found to be consistent regardless of the scale of each organization, which is measured by the average amount of tips received by each organization per year in both Model 2 and Model 3.
The findings implied that the potential temporal decay in dispatchers' emotional support might be generalizable and not limited to organizations with certain sizes and scales. 
Additionally, we also found that when the absolute value of users' polarity score is higher, as shown in Model 3 of Table \ref{Tab: emotional_support_Logistic}, which indicates that users are more likely to have negative emotions remaining by the end of conversations, dispatchers tend to provide more emotional support. 



Our analysis of conversation logs from a text-based safety incident reporting system revealed a decline in emotional support provided by dispatchers across different tip categories. However, we found that this decline was more pronounced for incidents that were more urgent, such as \textit{Emergency Messages}, \textit{Harassment / Abuse}, and \textit{Suspicious Activity}. 
On the other hand, this decline in emotional support was less noticeable for less-urgent events, such as theft or lost items. Additionally, our analysis showed that organizations with higher age/experience tend to provide less emotional support. While technology can be beneficial in terms of providing informational support (e.g., collecting information from users), our findings suggest that it may also lead to a degradation in emotional support from the first year of implementation.  
In an attempt to establish a systematic practice and positive word-of-mouth among users, organizations might incur an above-average level of service during the beginning of the implementation, while gradually reducing to a baseline during the years to come. 




\textbf{Summary (\textbf{RQ2})}: 
To understand what contextual factors were associated with the dispatchers' delivery of emotional support, we conducted a logistic regression analysis of dispatchers' messages.  
We identified factors related to emotional support delivery from three different aspects: tip-related factors, service-related factors, and organizational factors. 
From a tip-related perspective, dispatchers' likelihood to provide emotional support to users during reporting conversations was found to be associated with tip category, anonymity, and users' emotional polarity. For example, dispatchers tended to provide more emotional support for incidents under the categories of \textit{Harassment / Abuse} and \textit{Mental Health}, but less for incidents related to \textit{Drugs / Alcohol} and \textit{Noise Disturbance}.
From a service-related perspective, dispatchers were found to provide less emotional support during normal working hours during the day.
From an organizational perspective, a significant negative association was also found between the likelihood of providing emotional support and the number of years the organization had used the safety reporting system, suggesting a decline  of emotional support delivery after the initial period of system usage.

\subsection{DispatcherLLM Providing Informational and Emotional Support (RQ3)}
To investigate the potential of using large language models to improve the consistency of emotional support delivery, we introduced a domain-adapted LLM using the LiveSafe chat log dataset, and evaluated its effectiveness by comparing it with other LLMs.

\begin{table}[b]
\caption{\textit{dispatcherLLM} refers to our fine-tuned Llama 2 with retrieval-augmented dialogue generation. Our \textit{dispatcherLLM} outperforms the original Llama-2 and GPT-3.5 on the generated utterances across nine incident types that were used to train the model.  }

\begin{tabular*}{0.9\textwidth}{@{\extracolsep{\fill}}lll}
\hline
\multicolumn{3}{c}{\textbf{Comparison of models over all incidents}}                                                       \\  
\multicolumn{1}{l}{\textbf{Model}} & \multicolumn{1}{l}{\textbf{ROUGE}} & \multicolumn{1}{l}{\textbf{BERTScore}} \\ \hline

\textit{dispatcherLLM}              & \textbf{19.89}                     & \textbf{48.66}                          \\
Llama-2                                           & 6.20                       & 34.72                          \\
GPT-3.5                                               & 6.06                       & 40.09            \\

\hline      

\end{tabular*}

\label{tb:llm_overall}
\end{table}

\begin{figure}[t]
    \centering
    \includegraphics[width=.8\linewidth]{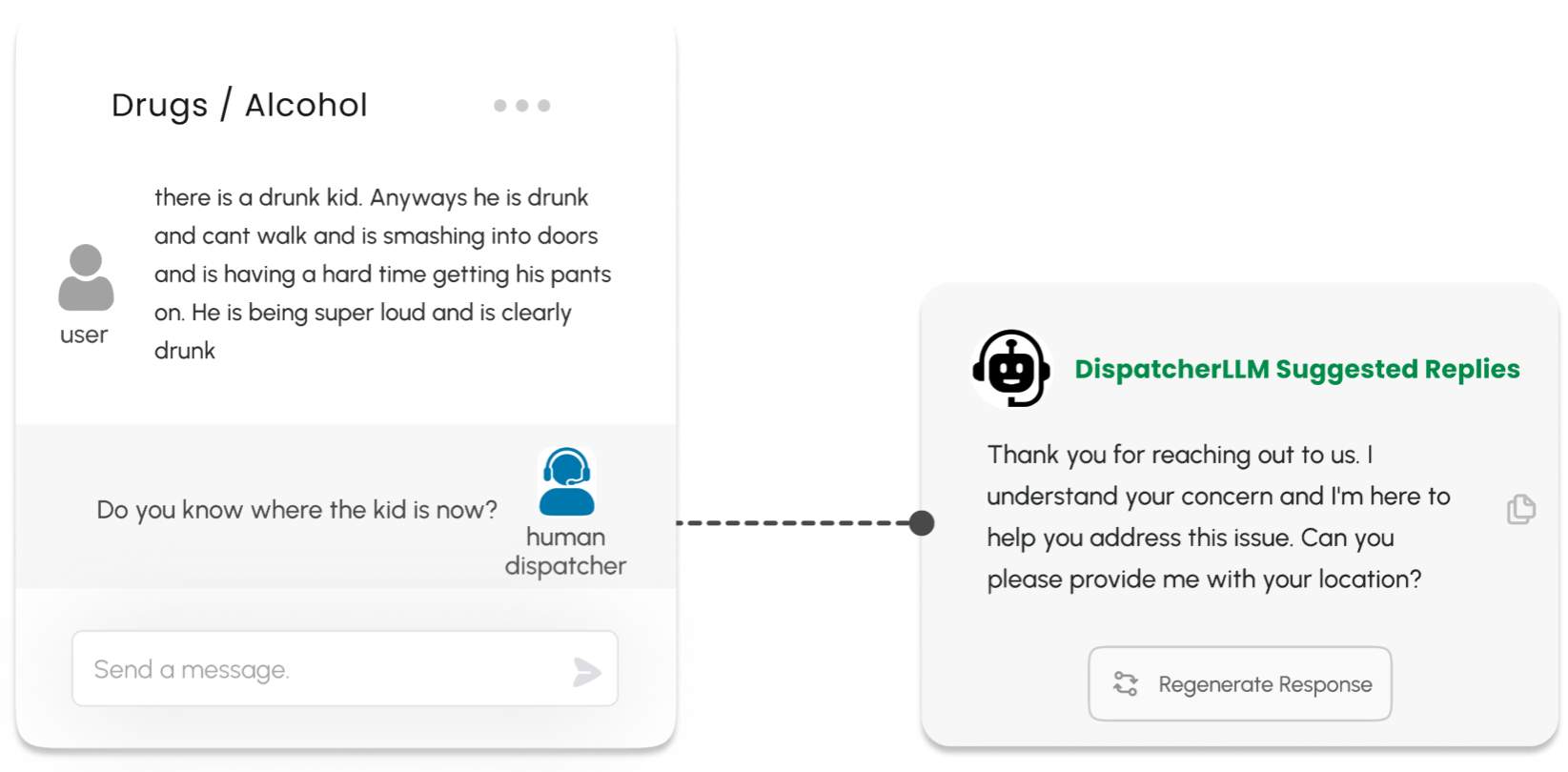}
    \caption{ Our \textit{dispatcherLLM} is used to automatically generate responses, given a user message. The left side of the figure shows an actual chat log between a user and a dispatcher, regarding a \textit{Drugs / Alcohol} incident. On the right, the figure displays a response generated by our \textit{dispatcherLLM}. This illustration serves not only to compare the two types of responses but also to demonstrate a potential use case scenario in which \textit{dispatcherLLM} suggests responses to human dispatchers, thereby enhancing the quality of their replies. 
    }
    \label{fig:LLM-case-1-sus}
\end{figure}
\subsubsection{Outperforming Popular LLMs for Overall Responses}
After training the \textit{dispatcherLLM} model following the method presented in section~\ref{sec:dispatcherLLM}, we first evaluated its performance in handling incident reports. 
For example, Fig.~\ref{fig:LLM-case-1-sus} shows a real chat between a user and a human dispatcher, when the user reported a suspicious activity. In the log, the human dispatcher only asked for location in a straightforward tone, whereas \textit{dispatcherLLM} provided emotional support (\textit{empathy display} according to \cite{liu2023discovering}) before asking for location information.
We then calculated the similarity between the suggested responses and those of human dispatchers' responses in the system log. We then compare these scores across 10,453 incidents against those generated by two popular LLMs, i.e., Llama-2 and GPT-3.5. Two commonly used metrics for evaluating language generation, ROUGE and BERTscore, \cite{zhang2020emoem, gao2021improving, minghui2022using} are used to show the performance. As shown in Table \ref{tb:llm_overall}, \textit{dispatcherLLM} yielded much better results than the other two models.

\subsubsection{Improved Emotional Support from \textit{dispatcherLLM}}
\label{sec:emotional_support_findings}

\begin{table}[!t]
\caption{Average emotional support rates for human dispatchers and \textit{dispatcherLLM} across different incident types. 
}
\centering
\begin{tabular}{lrccl}
\hline
\multicolumn{1}{c}{\textbf{Type}} & \multicolumn{1}{r}{\textbf{\#}} & \textbf{Human Dispatcher} & \textbf{\textit{dispatcherLLM}} & \textbf{\textit{p-value}} \\ \hline
Suspicious Activity              & 2,077        & 1.71                        & \textcolor{green}{\textbf{3.41}}             & $<0.001^{***}$            \\
Accident/Traffic/Parking         & 365         & \textbf{2.77}              & 2.22                      & $0.006^{**}$              \\
Drugs/Alcohol                    & 1,328        & 2.39                        & \textcolor{green}{\textbf{4.18}}             & $<0.001^{***}$            \\
Emergency Message                & 1,569        & 1.58                        & \textcolor{green}{\textbf{2.40}}             & $0.044^{*}$               \\
Facilities/Maintenance           & 756         & 7.23                        & 7.23                      & $0.004^{**}$              \\
Harassment/Abuse                 & 773         & 3.40                        & \textcolor{green}{\textbf{5.57}}             & $0.048^{*}$               \\
Mental Health                    & 272         & \textbf{5.66}             & 4.40                      & $0.012^{*}$               \\
Noise Disturbance                & 2,780        & 2.57                        & \textcolor{green}{\textbf{4.91}}             & $0.005^{**}$              \\
Theft/Lost Item                  & 533         & 7.53                        & 9.41             & 0.131                     \\ \hline
Total                            & 10,453       & 2.96                        & \textcolor{green}{\textbf{4.48}}             & $0.019^{*}$               \\ \hline
\end{tabular}

\label{tb:mean}
\end{table}

We further compared the emotional support rates between human dispatchers and \textit{dispatcherLLM}.  
We used the same method described in section \ref{sec:emotional_support_detection} to detect emotional support in the responses generated by \textit{dispatcherLLM}.  
The support rate was defined by the ratio of messages providing emotional support from human dispatchers or \textit{dispatcherLLM}. We compared the emotional support rates from two perspectives: incident category and time of service. 

First, we ran paired t-tests to assess the differences in emotional support between humans and \textit{dispatcherLLM}. 
Table \ref{tb:mean} presents the average support rate for different incident types.
The results show that \textit{dispatcherLLM} significantly increased emotional support in responses for six types of incidents, including: \textit{Suspicious Activity}, \textit{Drugs / Alcohol}, \textit{Emergency Message}, \textit{Harassment / Abuse}, \textit{Noise Disturbance}, and \textit{Theft / Lost Item}. 
On the other hand, human dispatchers showed a higher support rate in the \textit{Mental Health} category. 
Referring to the findings in RQ1 (Table~\ref{Tab:lr_Negativity}) and RQ2 (Table~\ref{Tab: emotional_support_Logistic}), it was observed that users did not express strong emotions in their messages when reporting \textit{Accident / Traffic / Parking} incidents. In contrast, users' emotional states were significantly more negative when reporting Mental Health incidents. This polarity in emotional states likely influenced the performance of \textit{dispatcherLLM}.

The second notable improvement concerned the temporal consistency is \textit{dispatcherLLM}'s delivery of emotional support. A Levene’s Test of Variances ($F=780.27, p<0.001^{***}$) showed that the LLM ($Z=0.51$) displayed a more consistent level of positive emotions throughout different hours of the day than humans ($Z=0.54$).
There is an interesting deficiency in emotional support provided by human dispatchers during daytime hours. Humans tend to become more emotionally supportive between 9:00 PM and 3:00 AM, suggesting a shift in their capacity for emotional engagement during nighttime hours. In comparison, \textit{dispatcherLLM}'s responses exhibited a more evenly distributed pattern of emotional positivity throughout the day. While \textit{dispatcherLLM}'s emotional support remained consistent, the human dispatcher's ability to offer emotional assistance experienced a distinct temporal variation, as shown in Table~\ref{Tab: emotional_support_Logistic}. This result holds practical value as it suggests that \textit{dispatcherLLM} potentially offers not only consistent availability but also maintains a stable emotional response, regardless of the time of service.

\subsubsection{\yirenrevised{Users' Perceptions of the Responses Generated by LLMs}}

\yirenrevised{To further understand how users perceive the emotional support in the responses generated by \textit{dispatcherLLM}, we conducted a survey study with human evaluators. }

\yirenrevised{During the study, the participants were given randomly sampled messages of incident reporting and asked to comment on whether emotional support should be provided in the example contexts and provide an explanation of their choice; if a participant responded with yes, they were further asked which LLM's (\textit{dispatcherLLM} or GPT-3.5) response could be used to improve the dispatcher's original reply, or if neither of them would be helpful. We also asked the participants to rate the appropriate level of emotional support. The major findings are highlighted below.}

\yirenrevised{First, participants thought \textit{dispatcherLLM} generated more human-like and \textit{``direct''} responses while GPT-3.5's responses were too lengthy (P10). 
We conducted a t-test showing a significant difference in the chat length between generated responses from human dispatchers and GPT-3.5 ($t = [-4.478], p = 0.011^{*}$), while no significant difference in chat length was identified between human dispatchers and \textit{dispatcherLLM}, which indicated a higher similarity in the linguistic style between human dispatcher and \textit{dispatcherLLM} than GPT-3.5.}

\yirenrevised{Second, participants agreed that in situations such as assault, abuse, and harassment, increased emotional support is essential. For example, Figure \ref{fig:survey_example} showed a random example incident of \textit{Harassment/Abuse}.
When rating whether the response provided an appropriate level of emotional support, participants gave \textit{dispatcherLLM} a significantly higher score over GPT-3.5 ($t = [2.135], p = 0.041^{*}$). They preferred the responses of \textit{dispatcherLLM}, because they found it provided both emotional support and \textit{``ensured that the appropriate actions are taken in a timely manner ...''} (P10). Several participants liked the conciseness of \textit{dispatcherLLM}'s response: \textit{``... it is concise and gives the appropriate amount of response with emotional support and call to action.''} (P7).
} %


\yirenrevised{Third, even when emotional support was not perceived as critical, participants still found the generated responses helpful in handling the incidents. For example, when reviewing a noise disturbance case, two-thirds of the participants thought ``a noise complaint probably doesn't require extensive emotional aid'' (P13). This comment aligns with the results shown in Table \ref{Tab:lr_Negativity}---where users expressed fewer negative emotions in their reports, and Table \ref{Tab: emotional_support_Logistic}---where dispatchers provided minimal emotional support. The lack of users' need for emotional support offered a potential explanation for why there was no significant difference in participants'  perceived level of support between \textit{dispatcherLLM} ($M=3.53, SD=0.94$) and GPT-3.5 ($M=3.35, SD=1.22$). Nonetheless, more than half of the participants thought the \textit{dispatcherLLM} responses would be helpful in handling the \textit{Noise Disturbance} incidents.}

\yirenrevised{Overall, participants' perspectives on how AI-provided emotional support varied across different types of incidents. More than 80\% of the participants agreed that AI needs to provide support in certain circumstances, particularly those involving mental health and emotional distress, e.g., ``anytime you sense distress in mental health scenarios'' (P1) and ``emergencies involving mental health or emotional crises need special care'' (P3). 
In a case where a mental health report was sent, 88.2\% participants preferred the generated response from GPT-3.5 ($t = [8.865], p < 0.001 ^{***}$).
As one participant explained, the \textit{``longer reply is showing that they truly do care and will do anything to support this person going through such a hard time''} (P1).
Considering the limited emotional support provided by human dispatchers during \textit{Mental Health} reporting, as shown in Table \ref{Tab: emotional_support_Logistic}, we may further improve LLM's emotional support by augmenting the training data from domain experts such as mental health counseling to fine-tune the \textit{dispatcherLLM} in future work, because the GPT-3.5 responses may lack professional expertise or not be sufficient to address the users' real need.} 



\begin{figure}[h!]
\resizebox{.95\textwidth}{!}{%
\begin{tabular}{p{.3\textwidth} p{.4\textwidth} p{.3\textwidth}}
\hline \hline
\textbf{User's First Message} &  & \\ \hline 

\parbox[t]{\textwidth}{\small My wife was reading at the Nobel library in a single room this afternoon. A guy stormed into her room and grabbed her hand. She was on the \#nd floor. She is really scared and shocked by this incident. That guy was laughing while in her room. He has curly hair and was wearing a hoody. She couldn't give any other information regarding his appearance. Please kindly look into it. Thank you.} 
& \\
\\
\hline
\hline
\textbf{Human Dispatcher} &  \textbf{GPT-3.5}&  \textbf{\textit{dispatcherLLM}}\\
\hline
\parbox[t]{.3\textwidth}{
\small Thank you for contacting the [ACRONYM] Police. Do you know how long ago this happened?\\
} & 
\parbox[t]{.4\textwidth}{
\small I'm so sorry to hear about this terrifying experience. Thank you for reaching out to us. It's important that we take immediate action to ensure your wife's safety and apprehend the individual responsible. I want to make sure your wife is safe right now. Can you please confirm her current location and provide me with any other pertinent details? Additionally, I want you to know that you can stay on the line with me so that I can offer support and guidance as we address this situation together. Let's work together to keep your wife safe and ensure that this individual is held accountable for their actions.\\ 
} & Thank you for reporting this incident to us. I'm so sorry to hear that your wife had this frightening experience. I'm here to help and support you through this. Can you please tell me more about the location of the library? Is it a public library? And do you know the time of the incident?
\\ \hline
\end{tabular}
}
\caption{
\yirenrevised{A randomly selected \textit{Harassment/Abuse} incident was showcased in the user survey. The follow-up question generated by \textit{dispatcherLLM} was perceived by participants to be more contextualized and specific than GPT-3.5's response.}
}
    \label{fig:survey_example}
\end{figure}

\textbf{Summary (\textbf{RQ3})}: 
The evaluation of \textit{dispatcherLLM} showcased its feasibility for assisting in the handling of safety incident reports. First, the generated responses of \textit{dispatcherLLM} were more similar to those of human dispatchers than those generated by the off-the-shelf model and GPT-3.5. The model was also observed to deliver greater emotional support than human dispatchers in two aspects: across various incident types and throughout different service hours. \yirenrevised{According to feedback from participants, they generally preferred \textit{dispatcherLLM}'s responses for they are more human-like, providing both emotional support and relevant instructions.  When emotional support is not deemed critical, participants still found the LLM-generated responses to be helpful in improving human dispatchers' incident report handling. It is important to note that their preferences for AI models also varied by incident type. In \textit{Mental Health} scenarios, participants favored longer and more empathetic responses. Future work should comprehensively evaluate these models in diverse contexts to offer more in-depth guidance. Human dispatchers should also be cautious about the over-reliance on the LLM-generated responses.} 

\section{Discussion}
In this section, we discuss the findings of this study and their connections to prior work and implications for future designs in the field of community safety risk reporting systems. 

\subsection{Temporal Dynamics of Emotional Support in Text-Based Safety Reporting}

\textbf{Users' negative emotions during incident reporting (RQ1)}
Existing studies have pointed out that users can exhibit negative emotions during traditional emergency incident reporting and can cause a negative impact on the handling process if not appropriately addressed by dispatchers \cite{feldman2021calming, Observations_on_the_Display}. 
The negativity in users' emotions, per our findings of \textbf{RQ1}, was also found in the reporting scenarios when using text-based reporting systems. A gradual decrease in both users' negative emotions and neutral utterances was found during conversations with dispatchers. This finding implied that two different types of support were provided by dispatchers to users during incident reporting sessions --- informational and emotional support, which has been discussed in prior research related to online chat-based healthcare consultation \cite{tan2020physician}. 
Within the context of text-based safety reporting, informational support refers to the process of both information collection and providing assistance, including dispatching officers, instructing, and providing additional information and resources. In terms of emotional support, the behavior of dispatchers accords with that of traditional voice-based incident reporting \cite{feldman2021calming}. 
The effect of dispatchers' emotional support delivery can be reflected through the changes in the negative emotions of users as dispatchers attempt to calm users.  

Our findings of \textbf{RQ1} analysis revealed an association of users' overall emotion with safety incidents' internal characteristics, i.e. tip categories. Incidents that are less severe or life-threatening in nature tend to trigger fewer negative emotions in users' responses. Another factor associated with users' emotional polarity is anonymity, as anonymous users tend to be more negatively polarized when making an incident report. 
This corresponds to findings of prior studies revealing anonymity, when offered as an option in online counseling systems \cite{gieselmann2016treating}, has the advantage of eliciting more truthful disclosure from users. 
However, no difference in users' emotional negativity was found across organizations with respect to the experience since the implementation of the safety reporting system and the number of tips received per year. This hints that users' initial emotion polarity only depends on the categories of incidents and users' choices in whether to report anonymously, as opposed to the characteristics of organizations. This indicates the consistency in levels of users' negativity to be expected by dispatchers regardless of their belonging organizations. On the other hand, higher learning institutions, which comprise the organizations where our analyzed data were collected, have a higher turnover rate of community members with the change caused by newly admitted students and graduates mostly in an annual cycle compared to other organizations and communities, which renders our findings to be more generalizable for communities with more frequent flows of members.

\textbf{Organizations' temporal decay in emotional support delivery (RQ2)} 
Although users exhibit consistent negative emotional displays across organizations, dispatchers were found to provide different levels of emotional support regarding how long organizations have implemented the system.
Temporally, organizations were found to have an elevated level of emotional support delivery during the first year of reporting system implementation. The ratio of provided emotional support gradually decreased to a baseline as organizations became more experienced in working with the system. This finding corresponds to a similar effect as ``initiative decay'' discussed in \citet{buchanan2005no}'s work.
This revealed two insights about the organizational adaptation of safety reporting systems: 
1) providing a high level of emotional support during the initial implementation of a safety reporting system may help to establish positive word-of-mouth and encourage ongoing use of the system;
2) as organizations become more experienced with using the system, the level of emotional support provided may decline, potentially due to a lack of awareness of the ongoing need for this type of support.
As the incentive for providing high-quality service with emotional support is reduced after the initial implementation period, it may be important for organizations to regularly assess the level of emotional support being provided to users of the safety reporting system, and to actively make efforts to maintain or increase this support as needed. This could involve providing training or resources for dispatchers to offer emotional support effectively, or implementing systems to monitor and track the level of support being provided. Ensuring that users continue to receive adequate emotional support can help to improve the overall effectiveness of the safety reporting system and increase user satisfaction.

Our findings also identified a difference in the nature of incidents reported by users by different time periods of the day. 
Dispatchers during certain time periods might suffer more from burnouts than others, depending on the distribution of incidents reported by users. For instance, incidents related to \textit{Noise Disturbance} and \textit{Drugs/Alcohol} were found to have a significant increase in numbers during the night shift (8 p.m. to 12 a.m.), which are often deemed to be more repetitive in terms of the procedures required to handle them. Additionally, more incidents related to \textit{Harassment/Abuse} were found to have been reported by community members during the night shift. This change in the composition of incident types caused an increase in both task volumes, repetitiveness and dispatchers' emotional cost required to handle the reports, leading to potentially higher emotional burnouts when compared to other shifts. This corresponds with our regression analysis results where even though the tip categories were controlled, there exists a persistent negative impact on emotional support delivery during the higher-volume shifts. 

\textbf{Improved emotional support from \textit{dispatcherLLM} (RQ3)}
\yirenrevised{
Prior studies have discussed how AI can be used to provide emotional support in various domains such as mental health counseling~\cite{denecke2020mental} and customer services~\cite{gelbrich2021emotional, hu2018touch}. Our study made the first attempt in the incident reporting domain by exploring the feasibility of using LLMs to provide both informational and emotional support. 
The findings from RQ3 demonstrated that LLM-based chat models are perceived as useful for improving responses from human dispatchers. 
According to the RQ2 results in Table \ref{Tab: emotional_support_Logistic}, \textit{dispatcherLLM} may be applied to support human dispatchers across various times of the day when they have psychological burnout \cite{tracy1998emotion, wang2022desired}. The AI's support can significantly enhance the consistency of the responses. 
For many incident categories, our developed \textit{dispatcherLLM} outperformed human dispatchers, except for \textit{Mental Health}. This underperformance might be partially attributed to the relative scarcity of training data in these specific categories. It is also possible that the human dispatcher chat logs used to fine-tune the \textit{dispatcherLLM} lack the necessary incident cases where deep expertise in emotional support is exemplified. Future studies should systematically evaluate LLMs' emotional support.  
Methods such as data augmentation \cite{Zheng2022AugESCDA, pan2023empathetic} and self-training \cite{sharma2021towards} can be utilized to further incorporate expertise from domains such as mental health counseling into the model. 
Similarly, the AI's support can help human dispatchers reflect more critically when handling diverse incidents. 
}

\subsection{Design Implications}


\textbf{Improving report handling with optimized workflow and AI agents} 
Our findings provided an example to support AI-enabled chatbots, as shown in Fig.~\ref{fig:LLM-case-1-sus}, where LLMs can be leveraged to assist human dispatchers with providing emotional support during safety-related incidents. This support may be particularly useful when a wide range of reports are submitted by a large amount of community members, because dispatchers may save time from typing by adopting part of the generated responses. Also, we found that both the volumes and categories among incidents reported by users differed across different time periods of a day, potentially causing varied pressure and burnouts for dispatchers on different work shifts \cite{Observations_on_the_Display}. Mixed reporting handling regardless of their nature in complexity and need for emotional support could lead to a degradation in dispatchers' service level in terms of emotional support delivery, as shown in the findings of \ref{sec:emo_support_categories}. 
For high-volume shifts where multiple dispatchers are present to provide support, dispatchers can be assigned dedicated categories of incidents based on the complexity and levels of emotional support needed. For example, playbooks and standard procedures for simpler and repetitive incidents can be created to increase handling efficiency \cite{aksin2007modern} so that a higher volume of incidents can be handled by a single dispatcher with a minimal negative impact on emotional support delivery, while incidents needing more attention for emotional support can be allocated to dispatchers with lower workload to mitigate burnouts. 
Certain categories of incidents, such as \textit{Noise Disturbance} and \textit{Theft / Lost Item}, with relatively less complex and repetitive procedures, can also benefit from partial automation using LLM-based agents. This can include the application of automatic information extraction \cite{minhas2022protecting} incorporating methods described in \ref{Event_argument_extraction_method} of this work, and contextualized dialogue response generation with emotional support similar to the model introduced in this study \cite{ma2020survey, liu2021towards, xu2022poke}. Designs should also consider assisting humans in providing emotional support without significantly increasing their cost of action, by offloading the handling of less complex incidents to LLMs. Methods such as automated response suggestions \cite{di2022idea} and escalation protocols for human intervention \cite{kamar2016directions} can be employed to facilitate a less intrusive human-AI collaboration process.
Nonetheless, how organizations integrate the proposed approach into their existing workflow requires longitudinal studies for thorough evaluation.

\textbf{Designing for anonymity in incident reporting}\label{anonymity}
In a safety incident reporting scenario, previous study \cite{ming2021examining} has suggested that the feature of allowing users to remain anonymous during reporting conversations plays a significant role in reducing users' concerns and helping users to provide more input to their submitted tips, which corresponds with the victimization theory \cite{lewis2017fear} as anonymity reduces users' perceived cost of reporting. 
Similar to the utility-cost model of reporting \cite{myers1980crimes}, dispatchers' decision-making process in providing emotional support can be conceptualized as a trade-off between the utility (e.g., resolving an urgent safety issue) and cost (e.g., time spent and emotional labor invoked \cite{tracy1998emotion, golding2017exploring}) of providing emotional support to users. 
Our finding implies the possibility that the display of users' anonymity status might potentially negatively impact dispatchers' evaluation of the validity of the reported incidents, which provides insights about improving the design of future reporting systems to better ensure the validity of reports and aligning users' and dispatchers' perception of anonymity. For example, safety administration might provide separate reporting channels for anonymous and non-anonymous reports with different levels of expected response time acknowledged by both stakeholders. 
Allowing asynchronous interaction between users and dispatchers leads to better communication and follow-up, compared to traditional communication channels (e.g., phone calls).

\section{Limitations and Future Work}
Although the findings in this study uncover unique insights about the role of emotions in safety reporting interaction between users and dispatchers, most analyses were conducted from a macro perspective. To further understand the effect of user emotion on dispatchers' decision-making process, more case-by-case qualitative analyses should be conducted to provide a deeper understanding of how the interactions took place. 
In the data pre-processing stage, we excluded any non-English tips. While this simplifies the analysis, it's possible that these tips could contain valuable information related to risk reporting. As a potential next step, analyses should be conducted from a multi-lingual perspective to examine the findings over speaking populations. 
\yirenrevised{During the evaluation of the \textit{dispatcherLLM} generated responses, community members served as surrogate human evaluators instead of actual victims,  which could limit the empirical value of the results. We are planning to run more comprehensive evaluation with professional dispatchers and actual victims by carefully addressing recruitment challenges and ethical concerns.} 
The study also provides implications for mitigating dispatchers' burnout, which is unable to be observed directly from the dataset itself. Further study should be conducted to provide more solid evidence about this inference, such as conducting interviews or surveys of the dispatcher population. 
When fine-tuning the LLM, all sampled chat logs were utilized for training and evaluation. Future work should further explore improving the data quality by filtering the chat logs based on criteria such as quality of emotional support to improve the model responses. 


\section{Conclusion}
In this study, we found that emotion plays a significant role in text-based reporting systems for community safety, with users' emotions varying significantly according to tip categories. 
Inconsistency in the emotional support provided by dispatchers was also identified, with organizations with more experience using the reporting system tending to provide less emotional support. 
We also piloted using LLM to facilitate safety incident report handling and identified its potential to provide better and more consistent emotional support.
Our findings offer new empirical evidence for understanding the patterns of users' emotions and organizations' emotional support from different perspectives and provide insights for the design of future text-based risk reporting systems using AI techniques including LLM.


\bibliographystyle{ACM-Reference-Format}
\bibliography{sources,livesafe}


\begin{thebibliography}{82}


\ifx \showCODEN    \undefined \def \showCODEN     #1{\unskip}     \fi
\ifx \showDOI      \undefined \def \showDOI       #1{#1}\fi
\ifx \showISBNx    \undefined \def \showISBNx     #1{\unskip}     \fi
\ifx \showISBNxiii \undefined \def \showISBNxiii  #1{\unskip}     \fi
\ifx \showISSN     \undefined \def \showISSN      #1{\unskip}     \fi
\ifx \showLCCN     \undefined \def \showLCCN      #1{\unskip}     \fi
\ifx \shownote     \undefined \def \shownote      #1{#1}          \fi
\ifx \showarticletitle \undefined \def \showarticletitle #1{#1}   \fi
\ifx \showURL      \undefined \def \showURL       {\relax}        \fi
\providecommand\bibfield[2]{#2}
\providecommand\bibinfo[2]{#2}
\providecommand\natexlab[1]{#1}
\providecommand\showeprint[2][]{arXiv:#2}

\bibitem[Ahmed et~al\mbox{.}(2014a)]%
        {Ahmed:2014:PPF:2611205.2557376}
\bibfield{author}{\bibinfo{person}{Syed~Ishtiaque Ahmed}, \bibinfo{person}{Steven~J. Jackson}, \bibinfo{person}{Nova Ahmed}, \bibinfo{person}{Hasan~Shahid Ferdous}, \bibinfo{person}{Md.~Rashidujjaman Rifat}, \bibinfo{person}{A.S.M Rizvi}, \bibinfo{person}{Shamir Ahmed}, {and} \bibinfo{person}{Rifat~Sabbir Mansur}.} \bibinfo{year}{2014}\natexlab{a}.
\newblock \showarticletitle{Protibadi: A Platform for Fighting Sexual Harassment in Urban Bangladesh}. In \bibinfo{booktitle}{\emph{Proceedings of the 32Nd Annual ACM Conference on Human Factors in Computing Systems}} (Toronto, Ontario, Canada) \emph{(\bibinfo{series}{CHI '14})}. \bibinfo{publisher}{ACM}, \bibinfo{address}{New York, NY, USA}, \bibinfo{pages}{2695--2704}.
\newblock
\showISBNx{978-1-4503-2473-1}
\urldef\tempurl%
\url{https://doi.org/10.1145/2556288.2557376}
\showDOI{\tempurl}


\bibitem[Ahmed et~al\mbox{.}(2014b)]%
        {ahmed2014protibadi}
\bibfield{author}{\bibinfo{person}{Syed~Ishtiaque Ahmed}, \bibinfo{person}{Steven~J Jackson}, \bibinfo{person}{Nova Ahmed}, \bibinfo{person}{Hasan~Shahid Ferdous}, \bibinfo{person}{Md~Rashidujjaman Rifat}, \bibinfo{person}{ASM Rizvi}, \bibinfo{person}{Shamir Ahmed}, {and} \bibinfo{person}{Rifat~Sabbir Mansur}.} \bibinfo{year}{2014}\natexlab{b}.
\newblock \showarticletitle{Protibadi: A platform for fighting sexual harassment in urban Bangladesh}. In \bibinfo{booktitle}{\emph{Proceedings of the SIGCHI Conference on Human Factors in Computing Systems}}. \bibinfo{pages}{2695--2704}.
\newblock


\bibitem[Aksin et~al\mbox{.}(2007)]%
        {aksin2007modern}
\bibfield{author}{\bibinfo{person}{Zeynep Aksin}, \bibinfo{person}{Mor Armony}, {and} \bibinfo{person}{Vijay Mehrotra}.} \bibinfo{year}{2007}\natexlab{}.
\newblock \showarticletitle{The modern call center: A multi-disciplinary perspective on operations management research}.
\newblock \bibinfo{journal}{\emph{Production and operations management}} \bibinfo{volume}{16}, \bibinfo{number}{6} (\bibinfo{year}{2007}), \bibinfo{pages}{665--688}.
\newblock


\bibitem[Ali et~al\mbox{.}(2023)]%
        {ali2023supporting}
\bibfield{author}{\bibinfo{person}{Farhan Ali}, \bibinfo{person}{Doris Choy}, \bibinfo{person}{Shanti Divaharan}, \bibinfo{person}{Hui~Yong Tay}, {and} \bibinfo{person}{Wenli Chen}.} \bibinfo{year}{2023}\natexlab{}.
\newblock \showarticletitle{Supporting self-directed learning and self-assessment using TeacherGAIA, a generative AI chatbot application: Learning approaches and prompt engineering}.
\newblock \bibinfo{journal}{\emph{Learning: Research and Practice}} \bibinfo{volume}{9}, \bibinfo{number}{2} (\bibinfo{year}{2023}), \bibinfo{pages}{135--147}.
\newblock


\bibitem[Barki et~al\mbox{.}(2007)]%
        {barki2007information}
\bibfield{author}{\bibinfo{person}{Henri Barki}, \bibinfo{person}{Ryad Titah}, {and} \bibinfo{person}{C{\'e}line Boffo}.} \bibinfo{year}{2007}\natexlab{}.
\newblock \showarticletitle{Information system use--related activity: an expanded behavioral conceptualization of individual-level information system use}.
\newblock \bibinfo{journal}{\emph{Information systems research}} \bibinfo{volume}{18}, \bibinfo{number}{2} (\bibinfo{year}{2007}), \bibinfo{pages}{173--192}.
\newblock


\bibitem[Bivens and Hasinoff(2018)]%
        {doi:10.1080/1369118X.2017.1309444}
\bibfield{author}{\bibinfo{person}{Rena Bivens} {and} \bibinfo{person}{Amy~Adele Hasinoff}.} \bibinfo{year}{2018}\natexlab{}.
\newblock \showarticletitle{Rape: is there an app for that? An empirical analysis of the features of anti-rape apps}.
\newblock \bibinfo{journal}{\emph{Information, Communication \& Society}} \bibinfo{volume}{21}, \bibinfo{number}{8} (\bibinfo{year}{2018}), \bibinfo{pages}{1050--1067}.
\newblock
\urldef\tempurl%
\url{https://doi.org/10.1080/1369118X.2017.1309444}
\showDOI{\tempurl}
\showeprint{https://doi.org/10.1080/1369118X.2017.1309444}


\bibitem[Blom et~al\mbox{.}(2010)]%
        {blom2010fear}
\bibfield{author}{\bibinfo{person}{Jan Blom}, \bibinfo{person}{Divya Viswanathan}, \bibinfo{person}{Mirjana Spasojevic}, \bibinfo{person}{Janet Go}, \bibinfo{person}{Karthik Acharya}, {and} \bibinfo{person}{Robert Ahonius}.} \bibinfo{year}{2010}\natexlab{}.
\newblock \showarticletitle{Fear and the city: role of mobile services in harnessing safety and security in urban use contexts}. In \bibinfo{booktitle}{\emph{Proceedings of the SIGCHI Conference on Human Factors in Computing Systems}}. \bibinfo{pages}{1841--1850}.
\newblock


\bibitem[Buchanan et~al\mbox{.}(2005)]%
        {buchanan2005no}
\bibfield{author}{\bibinfo{person}{David Buchanan}, \bibinfo{person}{Louise Fitzgerald}, \bibinfo{person}{Diane Ketley}, \bibinfo{person}{Rose Gollop}, \bibinfo{person}{Jane~Louise Jones}, \bibinfo{person}{Sharon~Saint Lamont}, \bibinfo{person}{Annette Neath}, {and} \bibinfo{person}{Elaine Whitby}.} \bibinfo{year}{2005}\natexlab{}.
\newblock \showarticletitle{No going back: A review of the literature on sustaining organizational change}.
\newblock \bibinfo{journal}{\emph{International Journal of Management Reviews}} \bibinfo{volume}{7}, \bibinfo{number}{3} (\bibinfo{year}{2005}), \bibinfo{pages}{189--205}.
\newblock


\bibitem[{CatapultEMS}(2022)]%
        {wetip}
\bibfield{author}{\bibinfo{person}{{CatapultEMS}}.} \bibinfo{year}{2022}\natexlab{}.
\newblock \bibinfo{title}{{WeTip Anonymous Reporting System}}.
\newblock \bibinfo{howpublished}{\url{https://www.wetip.com/}}.
\newblock
\newblock
\shownote{Online; accessed 06 July 2022}.


\bibitem[Chung et~al\mbox{.}(2023)]%
        {chung2023challenges}
\bibfield{author}{\bibinfo{person}{Neo~Christopher Chung}, \bibinfo{person}{George Dyer}, {and} \bibinfo{person}{Lennart Brocki}.} \bibinfo{year}{2023}\natexlab{}.
\newblock \showarticletitle{Challenges of large language models for mental health counseling}.
\newblock \bibinfo{journal}{\emph{arXiv preprint arXiv:2311.13857}} (\bibinfo{year}{2023}).
\newblock


\bibitem[Cvijikj et~al\mbox{.}(2015)]%
        {cvijikj2015towards}
\bibfield{author}{\bibinfo{person}{Irena~Pletikosa Cvijikj}, \bibinfo{person}{Cristina Kadar}, \bibinfo{person}{Bogdan Ivan}, {and} \bibinfo{person}{Yiea-Funk Te}.} \bibinfo{year}{2015}\natexlab{}.
\newblock \showarticletitle{Towards a crowdsourcing approach for crime prevention}. In \bibinfo{booktitle}{\emph{Adjunct Proceedings of the 2015 ACM International Joint Conference on Pervasive and Ubiquitous Computing and Proceedings of the 2015 ACM International Symposium on Wearable Computers}}. \bibinfo{pages}{1367--1372}.
\newblock


\bibitem[Daher et~al\mbox{.}(2020)]%
        {daher2020empathic}
\bibfield{author}{\bibinfo{person}{Karl Daher}, \bibinfo{person}{Jacky Casas}, \bibinfo{person}{Omar~Abou Khaled}, {and} \bibinfo{person}{Elena Mugellini}.} \bibinfo{year}{2020}\natexlab{}.
\newblock \showarticletitle{Empathic chatbot response for medical assistance}. In \bibinfo{booktitle}{\emph{Proceedings of the 20th ACM International Conference on Intelligent Virtual Agents}}. \bibinfo{pages}{1--3}.
\newblock


\bibitem[Demszky et~al\mbox{.}(2020)]%
        {demszky2020goemotions}
\bibfield{author}{\bibinfo{person}{Dorottya Demszky}, \bibinfo{person}{Dana Movshovitz-Attias}, \bibinfo{person}{Jeongwoo Ko}, \bibinfo{person}{Alan Cowen}, \bibinfo{person}{Gaurav Nemade}, {and} \bibinfo{person}{Sujith Ravi}.} \bibinfo{year}{2020}\natexlab{}.
\newblock \showarticletitle{GoEmotions: A dataset of fine-grained emotions}.
\newblock \bibinfo{journal}{\emph{arXiv preprint arXiv:2005.00547}} (\bibinfo{year}{2020}).
\newblock


\bibitem[Denecke et~al\mbox{.}(2020)]%
        {denecke2020mental}
\bibfield{author}{\bibinfo{person}{Kerstin Denecke}, \bibinfo{person}{Sayan Vaaheesan}, {and} \bibinfo{person}{Aaganya Arulnathan}.} \bibinfo{year}{2020}\natexlab{}.
\newblock \showarticletitle{A mental health chatbot for regulating emotions (SERMO)-concept and usability test}.
\newblock \bibinfo{journal}{\emph{IEEE Transactions on Emerging Topics in Computing}} \bibinfo{volume}{9}, \bibinfo{number}{3} (\bibinfo{year}{2020}), \bibinfo{pages}{1170--1182}.
\newblock


\bibitem[Di~Fede et~al\mbox{.}(2022)]%
        {di2022idea}
\bibfield{author}{\bibinfo{person}{Giulia Di~Fede}, \bibinfo{person}{Davide Rocchesso}, \bibinfo{person}{Steven~P Dow}, {and} \bibinfo{person}{Salvatore Andolina}.} \bibinfo{year}{2022}\natexlab{}.
\newblock \showarticletitle{The Idea Machine: LLM-based Expansion, Rewriting, Combination, and Suggestion of Ideas}. In \bibinfo{booktitle}{\emph{Proceedings of the 14th Conference on Creativity and Cognition}}. \bibinfo{pages}{623--627}.
\newblock


\bibitem[{Federal Communications Commission}(2022)]%
        {Textto911WhatYouNeedToKnow}
\bibfield{author}{\bibinfo{person}{{Federal Communications Commission}}.} \bibinfo{year}{2022}\natexlab{}.
\newblock \bibinfo{title}{{Text to 911: What You Need To Know}}.
\newblock \bibinfo{howpublished}{\url{https://www.fcc.gov/consumers/guides/what-you-need-know-about-text-911}}.
\newblock
\newblock
\shownote{Online; accessed 11 July 2022}.


\bibitem[Feldman(2021)]%
        {feldman2021calming}
\bibfield{author}{\bibinfo{person}{Heidi~Kevoe Feldman}.} \bibinfo{year}{2021}\natexlab{}.
\newblock \showarticletitle{Calming emotional 911 callers: Using redirection as a patient-focused directive in emergency medical calls}.
\newblock \bibinfo{journal}{\emph{Language \& Communication}}  \bibinfo{volume}{81} (\bibinfo{year}{2021}), \bibinfo{pages}{81--92}.
\newblock


\bibitem[Gao et~al\mbox{.}(2021)]%
        {gao2021improving}
\bibfield{author}{\bibinfo{person}{Jun Gao}, \bibinfo{person}{Yuhan Liu}, \bibinfo{person}{Haolin Deng}, \bibinfo{person}{Wei Wang}, \bibinfo{person}{Yu Cao}, \bibinfo{person}{Jiachen Du}, {and} \bibinfo{person}{Ruifeng Xu}.} \bibinfo{year}{2021}\natexlab{}.
\newblock \showarticletitle{Improving empathetic response generation by recognizing emotion cause in conversations}. In \bibinfo{booktitle}{\emph{Findings of the association for computational linguistics: EMNLP 2021}}. \bibinfo{pages}{807--819}.
\newblock


\bibitem[Gao et~al\mbox{.}(2020)]%
        {gao-machine-2020}
\bibfield{author}{\bibinfo{person}{Shuyang Gao}, \bibinfo{person}{Sanchit Agarwal}, \bibinfo{person}{Tagyoung Chung}, \bibinfo{person}{Di Jin}, {and} \bibinfo{person}{Dilek Hakkani{-}T{\"{u}}r}.} \bibinfo{year}{2020}\natexlab{}.
\newblock \showarticletitle{From Machine Reading Comprehension to Dialogue State Tracking: Bridging the Gap}.
\newblock \bibinfo{journal}{\emph{CoRR}}  \bibinfo{volume}{abs/2004.05827} (\bibinfo{year}{2020}).
\newblock
\showeprint[arXiv]{2004.05827}
\urldef\tempurl%
\url{https://arxiv.org/abs/2004.05827}
\showURL{%
\tempurl}


\bibitem[Gelbrich et~al\mbox{.}(2021)]%
        {gelbrich2021emotional}
\bibfield{author}{\bibinfo{person}{Katja Gelbrich}, \bibinfo{person}{Julia Hagel}, {and} \bibinfo{person}{Chiara Orsingher}.} \bibinfo{year}{2021}\natexlab{}.
\newblock \showarticletitle{Emotional support from a digital assistant in technology-mediated services: Effects on customer satisfaction and behavioral persistence}.
\newblock \bibinfo{journal}{\emph{International Journal of Research in Marketing}} \bibinfo{volume}{38}, \bibinfo{number}{1} (\bibinfo{year}{2021}), \bibinfo{pages}{176--193}.
\newblock


\bibitem[Ghandeharioun et~al\mbox{.}(2019)]%
        {ghandeharioun2019approximating}
\bibfield{author}{\bibinfo{person}{Asma Ghandeharioun}, \bibinfo{person}{Judy~Hanwen Shen}, \bibinfo{person}{Natasha Jaques}, \bibinfo{person}{Craig Ferguson}, \bibinfo{person}{Noah Jones}, \bibinfo{person}{Agata Lapedriza}, {and} \bibinfo{person}{Rosalind Picard}.} \bibinfo{year}{2019}\natexlab{}.
\newblock \showarticletitle{Approximating interactive human evaluation with self-play for open-domain dialog systems}.
\newblock \bibinfo{journal}{\emph{Advances in Neural Information Processing Systems}}  \bibinfo{volume}{32} (\bibinfo{year}{2019}).
\newblock


\bibitem[Gieselmann and Pietrowsky(2016)]%
        {gieselmann2016treating}
\bibfield{author}{\bibinfo{person}{Annika Gieselmann} {and} \bibinfo{person}{Reinhard Pietrowsky}.} \bibinfo{year}{2016}\natexlab{}.
\newblock \showarticletitle{Treating procrastination chat-based versus face-to-face: An RCT evaluating the role of self-disclosure and perceived counselor's characteristics}.
\newblock \bibinfo{journal}{\emph{Computers in Human Behavior}}  \bibinfo{volume}{54} (\bibinfo{year}{2016}), \bibinfo{pages}{444--452}.
\newblock


\bibitem[Golding et~al\mbox{.}(2017)]%
        {golding2017exploring}
\bibfield{author}{\bibinfo{person}{Sarah~E Golding}, \bibinfo{person}{Claire Horsfield}, \bibinfo{person}{Annette Davies}, \bibinfo{person}{Bernadette Egan}, \bibinfo{person}{Martyn Jones}, \bibinfo{person}{Mary Raleigh}, \bibinfo{person}{Patricia Schofield}, \bibinfo{person}{Allison Squires}, \bibinfo{person}{Kath Start}, \bibinfo{person}{Tom Quinn}, {et~al\mbox{.}}} \bibinfo{year}{2017}\natexlab{}.
\newblock \showarticletitle{Exploring the psychological health of emergency dispatch centre operatives: a systematic review and narrative synthesis}.
\newblock \bibinfo{journal}{\emph{PeerJ}}  \bibinfo{volume}{5} (\bibinfo{year}{2017}), \bibinfo{pages}{e3735}.
\newblock


\bibitem[Grace and Sinor(2021)]%
        {grace2021text}
\bibfield{author}{\bibinfo{person}{Rob Grace} {and} \bibinfo{person}{Sierra Sinor}.} \bibinfo{year}{2021}\natexlab{}.
\newblock \showarticletitle{How to text 911: A content analysis of text-to-911 public education information}. In \bibinfo{booktitle}{\emph{The 39th ACM International Conference on Design of Communication}}. \bibinfo{pages}{135--141}.
\newblock


\bibitem[Hall and Schmid~Mast(2007)]%
        {hall2007sources}
\bibfield{author}{\bibinfo{person}{Judith~A Hall} {and} \bibinfo{person}{Marianne Schmid~Mast}.} \bibinfo{year}{2007}\natexlab{}.
\newblock \showarticletitle{Sources of accuracy in the empathic accuracy paradigm.}
\newblock \bibinfo{journal}{\emph{Emotion}} \bibinfo{volume}{7}, \bibinfo{number}{2} (\bibinfo{year}{2007}), \bibinfo{pages}{438}.
\newblock


\bibitem[Hossain et~al\mbox{.}(2018)]%
        {hossain2018bangladesh}
\bibfield{author}{\bibinfo{person}{Md~Monsur Hossain}, \bibinfo{person}{Moushumi Sharmin}, {and} \bibinfo{person}{Shameem Ahmed}.} \bibinfo{year}{2018}\natexlab{}.
\newblock \showarticletitle{Bangladesh emergency services: a mobile application to provide 911-like service in Bangladesh}. In \bibinfo{booktitle}{\emph{Proceedings of the 1st ACM SIGCAS Conference on Computing and Sustainable Societies}}. \bibinfo{pages}{1--11}.
\newblock


\bibitem[Hu et~al\mbox{.}(2018)]%
        {hu2018touch}
\bibfield{author}{\bibinfo{person}{Tianran Hu}, \bibinfo{person}{Anbang Xu}, \bibinfo{person}{Zhe Liu}, \bibinfo{person}{Quanzeng You}, \bibinfo{person}{Yufan Guo}, \bibinfo{person}{Vibha Sinha}, \bibinfo{person}{Jiebo Luo}, {and} \bibinfo{person}{Rama Akkiraju}.} \bibinfo{year}{2018}\natexlab{}.
\newblock \showarticletitle{Touch your heart: A tone-aware chatbot for customer care on social media}. In \bibinfo{booktitle}{\emph{Proceedings of the 2018 CHI conference on human factors in computing systems}}. \bibinfo{pages}{1--12}.
\newblock


\bibitem[Huang et~al\mbox{.}(2023)]%
        {huang2023emotionally}
\bibfield{author}{\bibinfo{person}{Jen-tse Huang}, \bibinfo{person}{Man~Ho Lam}, \bibinfo{person}{Eric~John Li}, \bibinfo{person}{Shujie Ren}, \bibinfo{person}{Wenxuan Wang}, \bibinfo{person}{Wenxiang Jiao}, \bibinfo{person}{Zhaopeng Tu}, {and} \bibinfo{person}{Michael~R Lyu}.} \bibinfo{year}{2023}\natexlab{}.
\newblock \showarticletitle{Emotionally numb or empathetic? evaluating how llms feel using emotionbench}.
\newblock \bibinfo{journal}{\emph{arXiv preprint arXiv:2308.03656}} (\bibinfo{year}{2023}).
\newblock


\bibitem[Ignaco(2019)]%
        {ignaco2019development}
\bibfield{author}{\bibinfo{person}{Mary Ann~E Ignaco}.} \bibinfo{year}{2019}\natexlab{}.
\newblock \showarticletitle{Development of mobile application for incident reporting}. In \bibinfo{booktitle}{\emph{Proceedings of the 20th Annual SIG Conference on Information Technology Education}}. \bibinfo{pages}{162--162}.
\newblock


\bibitem[Iriberri et~al\mbox{.}(2006)]%
        {iriberri2006reporting}
\bibfield{author}{\bibinfo{person}{Alicia Iriberri}, \bibinfo{person}{Gondy Leroy}, {and} \bibinfo{person}{Nathan Garrett}.} \bibinfo{year}{2006}\natexlab{}.
\newblock \showarticletitle{Reporting on-campus crime online: User intention to use}. In \bibinfo{booktitle}{\emph{Proceedings of the 39th Annual Hawaii International Conference on System Sciences (HICSS'06)}}, Vol.~\bibinfo{volume}{4}. IEEE, \bibinfo{pages}{82a--82a}.
\newblock


\bibitem[Jo et~al\mbox{.}(2023)]%
        {jo2023understanding}
\bibfield{author}{\bibinfo{person}{Eunkyung Jo}, \bibinfo{person}{Daniel~A Epstein}, \bibinfo{person}{Hyunhoon Jung}, {and} \bibinfo{person}{Young-Ho Kim}.} \bibinfo{year}{2023}\natexlab{}.
\newblock \showarticletitle{Understanding the benefits and challenges of deploying conversational AI leveraging large language models for public health intervention}. In \bibinfo{booktitle}{\emph{Proceedings of the 2023 CHI Conference on Human Factors in Computing Systems}}. \bibinfo{pages}{1--16}.
\newblock


\bibitem[Kalyanchakravarthy et~al\mbox{.}(2014)]%
        {kalyanchakravarthy2014android}
\bibfield{author}{\bibinfo{person}{P Kalyanchakravarthy}, \bibinfo{person}{T Lakshmi}, \bibinfo{person}{R Rupavathi}, \bibinfo{person}{S Krishnadilip}, {and} \bibinfo{person}{P Lakshmankumar}.} \bibinfo{year}{2014}\natexlab{}.
\newblock \showarticletitle{Android Based Safety Triggering Application}.
\newblock \bibinfo{journal}{\emph{International Journal Of Computer Science and Information Technologies}} \bibinfo{volume}{5}, \bibinfo{number}{1} (\bibinfo{year}{2014}).
\newblock


\bibitem[Kamar(2016)]%
        {kamar2016directions}
\bibfield{author}{\bibinfo{person}{Ece Kamar}.} \bibinfo{year}{2016}\natexlab{}.
\newblock \showarticletitle{Directions in Hybrid Intelligence: Complementing AI Systems with Human Intelligence.}. In \bibinfo{booktitle}{\emph{IJCAI}}. \bibinfo{pages}{4070--4073}.
\newblock


\bibitem[Khanpour et~al\mbox{.}(2018)]%
        {khanpour2018identifying}
\bibfield{author}{\bibinfo{person}{Hamed Khanpour}, \bibinfo{person}{Cornelia Caragea}, {and} \bibinfo{person}{Prakhar Biyani}.} \bibinfo{year}{2018}\natexlab{}.
\newblock \showarticletitle{Identifying emotional support in online health communities}. In \bibinfo{booktitle}{\emph{Proceedings of the AAAI Conference on Artificial Intelligence}}, Vol.~\bibinfo{volume}{32}.
\newblock


\bibitem[Laestadius et~al\mbox{.}(2022)]%
        {laestadius2022too}
\bibfield{author}{\bibinfo{person}{Linnea Laestadius}, \bibinfo{person}{Andrea Bishop}, \bibinfo{person}{Michael Gonzalez}, \bibinfo{person}{Diana Illen{\v{c}}{\'\i}k}, {and} \bibinfo{person}{Celeste Campos-Castillo}.} \bibinfo{year}{2022}\natexlab{}.
\newblock \showarticletitle{Too human and not human enough: A grounded theory analysis of mental health harms from emotional dependence on the social chatbot Replika}.
\newblock \bibinfo{journal}{\emph{New Media \& Society}} (\bibinfo{year}{2022}), \bibinfo{pages}{14614448221142007}.
\newblock


\bibitem[Lewis(2017)]%
        {lewis2017fear}
\bibfield{author}{\bibinfo{person}{Dan~A Lewis}.} \bibinfo{year}{2017}\natexlab{}.
\newblock \bibinfo{booktitle}{\emph{Fear of crime: Incivility and the production of a social problem}}.
\newblock \bibinfo{publisher}{Routledge}.
\newblock


\bibitem[Li et~al\mbox{.}(2023)]%
        {li2023large}
\bibfield{author}{\bibinfo{person}{Cheng Li}, \bibinfo{person}{Jindong Wang}, \bibinfo{person}{Yixuan Zhang}, \bibinfo{person}{Kaijie Zhu}, \bibinfo{person}{Wenxin Hou}, \bibinfo{person}{Jianxun Lian}, \bibinfo{person}{Fang Luo}, \bibinfo{person}{Qiang Yang}, {and} \bibinfo{person}{Xing Xie}.} \bibinfo{year}{2023}\natexlab{}.
\newblock \showarticletitle{Large language models understand and can be enhanced by emotional stimuli}.
\newblock \bibinfo{journal}{\emph{arXiv preprint arXiv:2307.11760}} (\bibinfo{year}{2023}).
\newblock


\bibitem[Li et~al\mbox{.}(2021a)]%
        {li2021document}
\bibfield{author}{\bibinfo{person}{Sha Li}, \bibinfo{person}{Heng Ji}, {and} \bibinfo{person}{Jiawei Han}.} \bibinfo{year}{2021}\natexlab{a}.
\newblock \showarticletitle{Document-level event argument extraction by conditional generation}.
\newblock \bibinfo{journal}{\emph{arXiv preprint arXiv:2104.05919}} (\bibinfo{year}{2021}).
\newblock


\bibitem[Li et~al\mbox{.}(2021b)]%
        {li2021towards}
\bibfield{author}{\bibinfo{person}{Yanran Li}, \bibinfo{person}{Ke Li}, \bibinfo{person}{Hongke Ning}, \bibinfo{person}{Xiaoqiang Xia}, \bibinfo{person}{Yalong Guo}, \bibinfo{person}{Chen Wei}, \bibinfo{person}{Jianwei Cui}, {and} \bibinfo{person}{Bin Wang}.} \bibinfo{year}{2021}\natexlab{b}.
\newblock \showarticletitle{Towards an online empathetic chatbot with emotion causes}. In \bibinfo{booktitle}{\emph{Proceedings of the 44th international ACM SIGIR conference on research and development in information retrieval}}. \bibinfo{pages}{2041--2045}.
\newblock


\bibitem[Lin et~al\mbox{.}(2020)]%
        {lin2020joint}
\bibfield{author}{\bibinfo{person}{Ying Lin}, \bibinfo{person}{Heng Ji}, \bibinfo{person}{Fei Huang}, {and} \bibinfo{person}{Lingfei Wu}.} \bibinfo{year}{2020}\natexlab{}.
\newblock \showarticletitle{A joint neural model for information extraction with global features}. In \bibinfo{booktitle}{\emph{Proceedings of the 58th Annual Meeting of the Association for Computational Linguistics}}. \bibinfo{pages}{7999--8009}.
\newblock


\bibitem[Liu et~al\mbox{.}(2021)]%
        {liu2021towards}
\bibfield{author}{\bibinfo{person}{Siyang Liu}, \bibinfo{person}{Chujie Zheng}, \bibinfo{person}{Orianna Demasi}, \bibinfo{person}{Sahand Sabour}, \bibinfo{person}{Yu Li}, \bibinfo{person}{Zhou Yu}, \bibinfo{person}{Yong Jiang}, {and} \bibinfo{person}{Minlie Huang}.} \bibinfo{year}{2021}\natexlab{}.
\newblock \showarticletitle{Towards Emotional Support Dialog Systems}. In \bibinfo{booktitle}{\emph{Proceedings of the 59th Annual Meeting of the Association for Computational Linguistics and the 11th International Joint Conference on Natural Language Processing (Volume 1: Long Papers)}}. \bibinfo{pages}{3469--3483}.
\newblock


\bibitem[Liu et~al\mbox{.}(2023)]%
        {liu2023discovering}
\bibfield{author}{\bibinfo{person}{Yiren Liu}, \bibinfo{person}{Ryan Mayfield}, {and} \bibinfo{person}{Yun Huang}.} \bibinfo{year}{2023}\natexlab{}.
\newblock \showarticletitle{Discovering the Hidden Facts of User-Dispatcher Interactions via Text-based Reporting Systems for Community Safety}.
\newblock \bibinfo{journal}{\emph{Proceedings of the ACM on Human-Computer Interaction}} \bibinfo{volume}{7}, \bibinfo{number}{CSCW1} (\bibinfo{year}{2023}), \bibinfo{pages}{1--31}.
\newblock


\bibitem[Ma et~al\mbox{.}(2020)]%
        {ma2020survey}
\bibfield{author}{\bibinfo{person}{Yukun Ma}, \bibinfo{person}{Khanh~Linh Nguyen}, \bibinfo{person}{Frank~Z Xing}, {and} \bibinfo{person}{Erik Cambria}.} \bibinfo{year}{2020}\natexlab{}.
\newblock \showarticletitle{A survey on empathetic dialogue systems}.
\newblock \bibinfo{journal}{\emph{Information Fusion}}  \bibinfo{volume}{64} (\bibinfo{year}{2020}), \bibinfo{pages}{50--70}.
\newblock


\bibitem[Mendoza et~al\mbox{.}(2010)]%
        {mendoza2010software}
\bibfield{author}{\bibinfo{person}{Antonette Mendoza}, \bibinfo{person}{Jennie Carroll}, \bibinfo{person}{Linda Stern}, {et~al\mbox{.}}} \bibinfo{year}{2010}\natexlab{}.
\newblock \showarticletitle{Software appropriation over time: from adoption to stabilization and beyond}.
\newblock \bibinfo{journal}{\emph{Australasian Journal of Information Systems}} \bibinfo{volume}{16}, \bibinfo{number}{2} (\bibinfo{year}{2010}).
\newblock


\bibitem[Meng et~al\mbox{.}(2023)]%
        {meng2023mediated}
\bibfield{author}{\bibinfo{person}{Jingbo Meng}, \bibinfo{person}{Minjin Rheu}, \bibinfo{person}{Yue Zhang}, \bibinfo{person}{Yue Dai}, {and} \bibinfo{person}{Wei Peng}.} \bibinfo{year}{2023}\natexlab{}.
\newblock \showarticletitle{Mediated Social Support for Distress Reduction: AI Chatbots vs. Human}.
\newblock \bibinfo{journal}{\emph{Proceedings of the ACM on Human-Computer Interaction}} \bibinfo{volume}{7}, \bibinfo{number}{CSCW1} (\bibinfo{year}{2023}), \bibinfo{pages}{1--25}.
\newblock


\bibitem[Ming et~al\mbox{.}(2021)]%
        {ming2021examining}
\bibfield{author}{\bibinfo{person}{Shufan Ming}, \bibinfo{person}{Ryan~DW Mayfield}, \bibinfo{person}{Haocong Cheng}, \bibinfo{person}{Ke-Rou Wang}, {and} \bibinfo{person}{Yun Huang}.} \bibinfo{year}{2021}\natexlab{}.
\newblock \showarticletitle{Examining interactions between community members and university safety organizations through community-sourced risk systems}.
\newblock \bibinfo{journal}{\emph{Proceedings of the ACM on Human-Computer Interaction}} \bibinfo{volume}{5}, \bibinfo{number}{CSCW1} (\bibinfo{year}{2021}), \bibinfo{pages}{1--23}.
\newblock


\bibitem[Minghui et~al\mbox{.}(2022)]%
        {minghui2022using}
\bibfield{author}{\bibinfo{person}{Zou Minghui}, \bibinfo{person}{Pan Rui}, \bibinfo{person}{Zhang Sai}, {and} \bibinfo{person}{Zhang Xiaowang}.} \bibinfo{year}{2022}\natexlab{}.
\newblock \showarticletitle{Using Extracted Emotion Cause to Improve Content-Relevance for Empathetic Conversation Generation}. In \bibinfo{booktitle}{\emph{Proceedings of the 21st Chinese National Conference on Computational Linguistics}}. \bibinfo{pages}{811--823}.
\newblock


\bibitem[Minhas et~al\mbox{.}(2022)]%
        {minhas2022protecting}
\bibfield{author}{\bibinfo{person}{Rashid Minhas}, \bibinfo{person}{Camilla Elphick}, {and} \bibinfo{person}{Julia Shaw}.} \bibinfo{year}{2022}\natexlab{}.
\newblock \showarticletitle{Protecting victim and witness statement: examining the effectiveness of a chatbot that uses artificial intelligence and a cognitive interview}.
\newblock \bibinfo{journal}{\emph{AI \& SOCIETY}} \bibinfo{volume}{37}, \bibinfo{number}{1} (\bibinfo{year}{2022}), \bibinfo{pages}{265--281}.
\newblock


\bibitem[Myers(1980)]%
        {myers1980crimes}
\bibfield{author}{\bibinfo{person}{Samuel~L Myers}.} \bibinfo{year}{1980}\natexlab{}.
\newblock \showarticletitle{Why are Crimes Underreported? What is the Crime Rate? Does it" Really" Matter?}
\newblock \bibinfo{journal}{\emph{Social Science Quarterly}} \bibinfo{volume}{61}, \bibinfo{number}{1} (\bibinfo{year}{1980}), \bibinfo{pages}{23--43}.
\newblock


\bibitem[Pan et~al\mbox{.}(2023)]%
        {pan2023empathetic}
\bibfield{author}{\bibinfo{person}{Yanting Pan}, \bibinfo{person}{Yixuan Tang}, {and} \bibinfo{person}{Yuchen Niu}.} \bibinfo{year}{2023}\natexlab{}.
\newblock \showarticletitle{An Empathetic User-Centric Chatbot for Emotional Support}.
\newblock \bibinfo{journal}{\emph{arXiv preprint arXiv:2311.09271}} (\bibinfo{year}{2023}).
\newblock


\bibitem[Paoletti(2012)]%
        {paoletti2012operators}
\bibfield{author}{\bibinfo{person}{Isabella Paoletti}.} \bibinfo{year}{2012}\natexlab{}.
\newblock \showarticletitle{Operators managing callers’ sense of urgency in calls to the medical emergency number}.
\newblock \bibinfo{journal}{\emph{Pragmatics}} \bibinfo{volume}{22}, \bibinfo{number}{4} (\bibinfo{year}{2012}), \bibinfo{pages}{671--695}.
\newblock


\bibitem[Pinker and Bloom(1990)]%
        {pinker1990natural}
\bibfield{author}{\bibinfo{person}{Steven Pinker} {and} \bibinfo{person}{Paul Bloom}.} \bibinfo{year}{1990}\natexlab{}.
\newblock \showarticletitle{Natural language and natural selection}.
\newblock \bibinfo{journal}{\emph{Behavioral and brain sciences}} \bibinfo{volume}{13}, \bibinfo{number}{4} (\bibinfo{year}{1990}), \bibinfo{pages}{707--727}.
\newblock


\bibitem[Reaves(2015)]%
        {reaves2015campus}
\bibfield{author}{\bibinfo{person}{Brian~A Reaves}.} \bibinfo{year}{2015}\natexlab{}.
\newblock \bibinfo{booktitle}{\emph{Campus law enforcement, 2011-12}}.
\newblock \bibinfo{publisher}{US Department of Justice, Office of Justice Programs, Bureau of Justice}.
\newblock


\bibitem[Sakpere et~al\mbox{.}(2015)]%
        {sakpere2015usable}
\bibfield{author}{\bibinfo{person}{Aderonke~Busayo Sakpere}, \bibinfo{person}{Anne~VDM Kayem}, {and} \bibinfo{person}{Thabo Ndlovu}.} \bibinfo{year}{2015}\natexlab{}.
\newblock \showarticletitle{A usable and secure crime reporting system for technology resource constrained context}. In \bibinfo{booktitle}{\emph{2015 IEEE 29th International Conference on Advanced Information Networking and Applications Workshops}}. IEEE, \bibinfo{pages}{424--429}.
\newblock


\bibitem[Scherer et~al\mbox{.}(2001)]%
        {scherer2001emotion}
\bibfield{author}{\bibinfo{person}{Klaus~R Scherer}, \bibinfo{person}{Rainer Banse}, {and} \bibinfo{person}{Harald~G Wallbott}.} \bibinfo{year}{2001}\natexlab{}.
\newblock \showarticletitle{Emotion inferences from vocal expression correlate across languages and cultures}.
\newblock \bibinfo{journal}{\emph{Journal of Cross-cultural psychology}} \bibinfo{volume}{32}, \bibinfo{number}{1} (\bibinfo{year}{2001}), \bibinfo{pages}{76--92}.
\newblock


\bibitem[Sharma et~al\mbox{.}(2021)]%
        {sharma2021towards}
\bibfield{author}{\bibinfo{person}{Ashish Sharma}, \bibinfo{person}{Inna~W Lin}, \bibinfo{person}{Adam~S Miner}, \bibinfo{person}{David~C Atkins}, {and} \bibinfo{person}{Tim Althoff}.} \bibinfo{year}{2021}\natexlab{}.
\newblock \showarticletitle{Towards facilitating empathic conversations in online mental health support: A reinforcement learning approach}. In \bibinfo{booktitle}{\emph{Proceedings of the Web Conference 2021}}. \bibinfo{pages}{194--205}.
\newblock


\bibitem[Shuyo(2010)]%
        {nakatani2010langdetect}
\bibfield{author}{\bibinfo{person}{Nakatani Shuyo}.} \bibinfo{year}{2010}\natexlab{}.
\newblock \bibinfo{title}{Language Detection Library for Java}.
\newblock
\newblock
\urldef\tempurl%
\url{http://code.google.com/p/language-detection/}
\showURL{%
\tempurl}


\bibitem[Sijbrandij et~al\mbox{.}(2020)]%
        {sijbrandij2020effect}
\bibfield{author}{\bibinfo{person}{Marit Sijbrandij}, \bibinfo{person}{Rebecca Horn}, \bibinfo{person}{Rebecca Esliker}, \bibinfo{person}{Fiona O’may}, \bibinfo{person}{Relinde Reiffers}, \bibinfo{person}{Leontien Ruttenberg}, \bibinfo{person}{Kimberly Stam}, \bibinfo{person}{Joop de Jong}, {and} \bibinfo{person}{Alastair Ager}.} \bibinfo{year}{2020}\natexlab{}.
\newblock \showarticletitle{The effect of psychological first aid training on knowledge and understanding about psychosocial support principles: a cluster-randomized controlled trial}.
\newblock \bibinfo{journal}{\emph{International journal of environmental research and public health}} \bibinfo{volume}{17}, \bibinfo{number}{2} (\bibinfo{year}{2020}), \bibinfo{pages}{484}.
\newblock


\bibitem[Siv{\v{c}}evi{\'c} et~al\mbox{.}(2020)]%
        {sivvcevic2020possibilities}
\bibfield{author}{\bibinfo{person}{Du{\v{s}}ko Siv{\v{c}}evi{\'c}}, \bibinfo{person}{Ivan Ko{\v{s}}anin}, \bibinfo{person}{Slobodan Nedeljkovi{\'c}}, \bibinfo{person}{Vojkan Nikoli{\'c}}, \bibinfo{person}{Kristijan Kuk}, {and} \bibinfo{person}{Sr{\dj}an Nogo}.} \bibinfo{year}{2020}\natexlab{}.
\newblock \showarticletitle{Possibilities of used intelligence based agents in instant messaging on e-government services}. In \bibinfo{booktitle}{\emph{2020 19th International Symposium INFOTEH-JAHORINA (INFOTEH)}}. IEEE, \bibinfo{pages}{1--5}.
\newblock


\bibitem[Tan and Yan(2020)]%
        {tan2020physician}
\bibfield{author}{\bibinfo{person}{Hongying Tan} {and} \bibinfo{person}{Mengling Yan}.} \bibinfo{year}{2020}\natexlab{}.
\newblock \showarticletitle{Physician-user interaction and users' perceived service quality: evidence from Chinese mobile healthcare consultation}.
\newblock \bibinfo{journal}{\emph{Information Technology \& People}} (\bibinfo{year}{2020}).
\newblock


\bibitem[Touvron et~al\mbox{.}(2023)]%
        {touvron2023llama}
\bibfield{author}{\bibinfo{person}{Hugo Touvron}, \bibinfo{person}{Louis Martin}, \bibinfo{person}{Kevin Stone}, \bibinfo{person}{Peter Albert}, \bibinfo{person}{Amjad Almahairi}, \bibinfo{person}{Yasmine Babaei}, \bibinfo{person}{Nikolay Bashlykov}, \bibinfo{person}{Soumya Batra}, \bibinfo{person}{Prajjwal Bhargava}, \bibinfo{person}{Shruti Bhosale}, {et~al\mbox{.}}} \bibinfo{year}{2023}\natexlab{}.
\newblock \showarticletitle{Llama 2: Open foundation and fine-tuned chat models}.
\newblock \bibinfo{journal}{\emph{arXiv preprint arXiv:2307.09288}} (\bibinfo{year}{2023}).
\newblock


\bibitem[Tracy(2002)]%
        {tracy2002questioning}
\bibfield{author}{\bibinfo{person}{Sarah~J Tracy}.} \bibinfo{year}{2002}\natexlab{}.
\newblock \showarticletitle{When questioning turns to face threat: An interactional sensitivity in 911 call-taking}.
\newblock \bibinfo{journal}{\emph{Western Journal of Communication (includes Communication Reports)}} \bibinfo{volume}{66}, \bibinfo{number}{2} (\bibinfo{year}{2002}), \bibinfo{pages}{129--157}.
\newblock


\bibitem[Tracy and Tracy(1998)]%
        {tracy1998emotion}
\bibfield{author}{\bibinfo{person}{Sarah~J Tracy} {and} \bibinfo{person}{Karen Tracy}.} \bibinfo{year}{1998}\natexlab{}.
\newblock \showarticletitle{Emotion labor at 911: A case study and theoretical critique}.
\newblock  (\bibinfo{year}{1998}).
\newblock


\bibitem[Tyre and Orlikowski(1994)]%
        {tyre1994windows}
\bibfield{author}{\bibinfo{person}{Marcie~J Tyre} {and} \bibinfo{person}{Wanda~J Orlikowski}.} \bibinfo{year}{1994}\natexlab{}.
\newblock \showarticletitle{Windows of opportunity: Temporal patterns of technological adaptation in organizations}.
\newblock \bibinfo{journal}{\emph{Organization science}} \bibinfo{volume}{5}, \bibinfo{number}{1} (\bibinfo{year}{1994}), \bibinfo{pages}{98--118}.
\newblock


\bibitem[{University of Illinois Chicago}(2022)]%
        {uicSAFE}
\bibfield{author}{\bibinfo{person}{{University of Illinois Chicago}}.} \bibinfo{year}{2022}\natexlab{}.
\newblock \bibinfo{title}{{UIC SAFE App Office of Preparedness and Response University of Illinois Chicago}}.
\newblock \bibinfo{howpublished}{\url{https://ready.uic.edu/toolkit/uic-safe-app/}}.
\newblock
\newblock
\shownote{Online; accessed 06 July 2022}.


\bibitem[Vaux et~al\mbox{.}(1987)]%
        {vaux1987modes}
\bibfield{author}{\bibinfo{person}{Alan Vaux}, \bibinfo{person}{Sharon Riedel}, {and} \bibinfo{person}{Doreen Stewart}.} \bibinfo{year}{1987}\natexlab{}.
\newblock \showarticletitle{Modes of social support: The social support behaviors (SS-B) scale}.
\newblock \bibinfo{journal}{\emph{American Journal of Community Psychology}} \bibinfo{volume}{15}, \bibinfo{number}{2} (\bibinfo{year}{1987}), \bibinfo{pages}{209--232}.
\newblock


\bibitem[Walker et~al\mbox{.}(2006)]%
        {walker2006ace}
\bibfield{author}{\bibinfo{person}{Christopher Walker}, \bibinfo{person}{Stephanie Strassel}, \bibinfo{person}{Julie Medero}, {and} \bibinfo{person}{Kazuaki Maeda}.} \bibinfo{year}{2006}\natexlab{}.
\newblock \showarticletitle{ACE 2005 multilingual training corpus}.
\newblock \bibinfo{journal}{\emph{Linguistic Data Consortium, Philadelphia}}  \bibinfo{volume}{57} (\bibinfo{year}{2006}), \bibinfo{pages}{45}.
\newblock


\bibitem[Wang et~al\mbox{.}(2021)]%
        {wang2021cass}
\bibfield{author}{\bibinfo{person}{Liuping Wang}, \bibinfo{person}{Dakuo Wang}, \bibinfo{person}{Feng Tian}, \bibinfo{person}{Zhenhui Peng}, \bibinfo{person}{Xiangmin Fan}, \bibinfo{person}{Zhan Zhang}, \bibinfo{person}{Mo Yu}, \bibinfo{person}{Xiaojuan Ma}, {and} \bibinfo{person}{Hongan Wang}.} \bibinfo{year}{2021}\natexlab{}.
\newblock \showarticletitle{Cass: Towards building a social-support chatbot for online health community}.
\newblock \bibinfo{journal}{\emph{Proceedings of the ACM on Human-Computer Interaction}} \bibinfo{volume}{5}, \bibinfo{number}{CSCW1} (\bibinfo{year}{2021}), \bibinfo{pages}{1--31}.
\newblock


\bibitem[Wang(2022)]%
        {wang2022desired}
\bibfield{author}{\bibinfo{person}{Phoenix~Chi Wang}.} \bibinfo{year}{2022}\natexlab{}.
\newblock \showarticletitle{The ‘desired state of mind’: Emotional labor and the hidden cost of symbolic power in 911 emergency response in the US}.
\newblock \bibinfo{journal}{\emph{Ethnography}} (\bibinfo{year}{2022}), \bibinfo{pages}{14661381221145353}.
\newblock


\bibitem[Wang et~al\mbox{.}(2023)]%
        {wang2023emotional}
\bibfield{author}{\bibinfo{person}{Xuena Wang}, \bibinfo{person}{Xueting Li}, \bibinfo{person}{Zi Yin}, \bibinfo{person}{Yue Wu}, {and} \bibinfo{person}{Jia Liu}.} \bibinfo{year}{2023}\natexlab{}.
\newblock \showarticletitle{Emotional intelligence of large language models}.
\newblock \bibinfo{journal}{\emph{Journal of Pacific Rim Psychology}}  \bibinfo{volume}{17} (\bibinfo{year}{2023}), \bibinfo{pages}{18344909231213958}.
\newblock


\bibitem[Weber et~al\mbox{.}(2004)]%
        {weber2004communcating}
\bibfield{author}{\bibinfo{person}{Keith Weber}, \bibinfo{person}{Aaron Johnson}, {and} \bibinfo{person}{Michael Corrigan}.} \bibinfo{year}{2004}\natexlab{}.
\newblock \showarticletitle{Communcating emotional support and its relationship to feelings of being understood, trust, and self-disclosure}.
\newblock \bibinfo{journal}{\emph{Communication research reports}} \bibinfo{volume}{21}, \bibinfo{number}{3} (\bibinfo{year}{2004}), \bibinfo{pages}{316--323}.
\newblock


\bibitem[Whalen and Zimmerman(1998)]%
        {Observations_on_the_Display}
\bibfield{author}{\bibinfo{person}{Jack Whalen} {and} \bibinfo{person}{Don~H. Zimmerman}.} \bibinfo{year}{1998}\natexlab{}.
\newblock \showarticletitle{Observations on the Display and Management of Emotion in Naturally Occurring Activities: The Case of "Hysteria" in Calls to 9-1-1}.
\newblock \bibinfo{journal}{\emph{Social Psychology Quarterly}} \bibinfo{volume}{61}, \bibinfo{number}{2} (\bibinfo{year}{1998}), \bibinfo{pages}{141--159}.
\newblock
\showISSN{01902725}
\urldef\tempurl%
\url{http://www.jstor.org/stable/2787066}
\showURL{%
\tempurl}


\bibitem[Xia et~al\mbox{.}(2019)]%
        {xia2019multi}
\bibfield{author}{\bibinfo{person}{Congying Xia}, \bibinfo{person}{Chenwei Zhang}, \bibinfo{person}{Tao Yang}, \bibinfo{person}{Yaliang Li}, \bibinfo{person}{Nan Du}, \bibinfo{person}{Xian Wu}, \bibinfo{person}{Wei Fan}, \bibinfo{person}{Fenglong Ma}, {and} \bibinfo{person}{Philip Yu}.} \bibinfo{year}{2019}\natexlab{}.
\newblock \showarticletitle{Multi-grained named entity recognition}.
\newblock \bibinfo{journal}{\emph{arXiv preprint arXiv:1906.08449}} (\bibinfo{year}{2019}).
\newblock


\bibitem[Xu et~al\mbox{.}(2017)]%
        {xu2017new}
\bibfield{author}{\bibinfo{person}{Anbang Xu}, \bibinfo{person}{Zhe Liu}, \bibinfo{person}{Yufan Guo}, \bibinfo{person}{Vibha Sinha}, {and} \bibinfo{person}{Rama Akkiraju}.} \bibinfo{year}{2017}\natexlab{}.
\newblock \showarticletitle{A new chatbot for customer service on social media}. In \bibinfo{booktitle}{\emph{Proceedings of the 2017 CHI conference on human factors in computing systems}}. \bibinfo{pages}{3506--3510}.
\newblock


\bibitem[Xu et~al\mbox{.}(2022)]%
        {xu2022poke}
\bibfield{author}{\bibinfo{person}{Xiaohan Xu}, \bibinfo{person}{Xuying Meng}, {and} \bibinfo{person}{Yequan Wang}.} \bibinfo{year}{2022}\natexlab{}.
\newblock \showarticletitle{PoKE: Prior Knowledge Enhanced Emotional Support Conversation with Latent Variable}.
\newblock \bibinfo{journal}{\emph{arXiv preprint arXiv:2210.12640}} (\bibinfo{year}{2022}).
\newblock


\bibitem[Yabe(2017)]%
        {yabe2017cost}
\bibfield{author}{\bibinfo{person}{Manako Yabe}.} \bibinfo{year}{2017}\natexlab{}.
\newblock \showarticletitle{Cost-benefit evaluation: Students’, faculty’s, and staff’s willingness to pay for a campus safety app}.
\newblock \bibinfo{journal}{\emph{Journal of Criminal Justice Education}} \bibinfo{volume}{28}, \bibinfo{number}{2} (\bibinfo{year}{2017}), \bibinfo{pages}{207--221}.
\newblock


\bibitem[Zhang et~al\mbox{.}(2020)]%
        {zhang2020emoem}
\bibfield{author}{\bibinfo{person}{Ao Zhang}, \bibinfo{person}{Shaojuan Wu}, \bibinfo{person}{Xiaowang Zhang}, \bibinfo{person}{Shizhan Chen}, \bibinfo{person}{Yuchun Shu}, {and} \bibinfo{person}{Zhiyong Feng}.} \bibinfo{year}{2020}\natexlab{}.
\newblock \showarticletitle{EmoEM: Emotional Expression in a Multi-turn Dialogue Model}. In \bibinfo{booktitle}{\emph{2020 IEEE 32nd International Conference on Tools with Artificial Intelligence (ICTAI)}}. IEEE, \bibinfo{pages}{496--501}.
\newblock


\bibitem[Zhang et~al\mbox{.}(2009)]%
        {zhang2009sentiment}
\bibfield{author}{\bibinfo{person}{Changli Zhang}, \bibinfo{person}{Daniel Zeng}, \bibinfo{person}{Jiexun Li}, \bibinfo{person}{Fei-Yue Wang}, {and} \bibinfo{person}{Wanli Zuo}.} \bibinfo{year}{2009}\natexlab{}.
\newblock \showarticletitle{Sentiment analysis of Chinese documents: From sentence to document level}.
\newblock \bibinfo{journal}{\emph{Journal of the American Society for Information Science and Technology}} \bibinfo{volume}{60}, \bibinfo{number}{12} (\bibinfo{year}{2009}), \bibinfo{pages}{2474--2487}.
\newblock


\bibitem[Zheng et~al\mbox{.}(2022)]%
        {Zheng2022AugESCDA}
\bibfield{author}{\bibinfo{person}{Chujie Zheng}, \bibinfo{person}{Sahand Sabour}, \bibinfo{person}{Jiaxin Wen}, \bibinfo{person}{Zheng Zhang}, {and} \bibinfo{person}{Minlie Huang}.} \bibinfo{year}{2022}\natexlab{}.
\newblock \showarticletitle{AugESC: Dialogue Augmentation with Large Language Models for Emotional Support Conversation}. In \bibinfo{booktitle}{\emph{Annual Meeting of the Association for Computational Linguistics}}.
\newblock
\urldef\tempurl%
\url{https://api.semanticscholar.org/CorpusID:258588110}
\showURL{%
\tempurl}


\bibitem[Zheng et~al\mbox{.}(2023b)]%
        {zheng2023augesc}
\bibfield{author}{\bibinfo{person}{Chujie Zheng}, \bibinfo{person}{Sahand Sabour}, \bibinfo{person}{Jiaxin Wen}, \bibinfo{person}{Zheng Zhang}, {and} \bibinfo{person}{Minlie Huang}.} \bibinfo{year}{2023}\natexlab{b}.
\newblock \showarticletitle{Augesc: Dialogue augmentation with large language models for emotional support conversation}. In \bibinfo{booktitle}{\emph{Findings of the Association for Computational Linguistics: ACL 2023}}. \bibinfo{pages}{1552--1568}.
\newblock


\bibitem[Zheng et~al\mbox{.}(2023c)]%
        {zheng2023synergizing}
\bibfield{author}{\bibinfo{person}{Qingxiao Zheng}, \bibinfo{person}{Zhongwei Xu}, \bibinfo{person}{Abhinav Choudhary}, \bibinfo{person}{Yuting Chen}, \bibinfo{person}{Yongming Li}, {and} \bibinfo{person}{Yun Huang}.} \bibinfo{year}{2023}\natexlab{c}.
\newblock \showarticletitle{Synergizing human-AI agency: a guide of 23 heuristics for service co-creation with LLM-based agents}.
\newblock \bibinfo{journal}{\emph{arXiv preprint arXiv:2310.15065}} (\bibinfo{year}{2023}).
\newblock


\bibitem[Zheng et~al\mbox{.}(2023a)]%
        {zheng2023building}
\bibfield{author}{\bibinfo{person}{Zhonghua Zheng}, \bibinfo{person}{Lizi Liao}, \bibinfo{person}{Yang Deng}, {and} \bibinfo{person}{Liqiang Nie}.} \bibinfo{year}{2023}\natexlab{a}.
\newblock \showarticletitle{Building emotional support chatbots in the era of llms}.
\newblock \bibinfo{journal}{\emph{arXiv preprint arXiv:2308.11584}} (\bibinfo{year}{2023}).
\newblock


\end{thebibliography}

\clearpage
\appendix
\onecolumn
\section{APPENDIX}
\label{appendix}
\subsection{Definitions of the Ontology}

\begin{table}[h!]
\small
\resizebox{\linewidth}{!}{
\begin{tabular}{llp{10cm}}
\hline \hline
\multicolumn{3}{c}{\textbf{Definition of Event Entities}}                                                                                                                       \\ \hline \hline
\multicolumn{1}{l|}{\textbf{Entity}}              & \multicolumn{1}{l|}{\textbf{SubEntity}}   & \textbf{Definition}                                                              \\ \hline
\multicolumn{1}{l|}{Person}                       & \multicolumn{1}{l|}{---}                    & A single person or a group of people                                             \\ \hline
\multicolumn{1}{l|}{Location}                     & \multicolumn{1}{l|}{---}                    & A location denoted as a point such as in a postal system or abstract coordinates \\ \hline
\multicolumn{1}{l|}{Weapon}                       & \multicolumn{1}{l|}{---}                    & The primary method/instrument used by the offender which causes harm             \\ \hline
\multicolumn{1}{l|}{Time}                         & \multicolumn{1}{l|}{---}                    & The time an incident occurred                                                    \\ \hline
\multicolumn{1}{l|}{Object}                       & \multicolumn{1}{l|}{---}                    & An inanimate object involved in the incident                                     \\ \hline
\multicolumn{1}{l|}{\multirow{9}{*}{Description}} & \multicolumn{1}{l|}{Person-Age}           & The age of the person                                                            \\ \cline{2-3} 
\multicolumn{1}{l|}{}                             & \multicolumn{1}{l|}{Person-Race}          & The racial description of the person                                             \\ \cline{2-3} 
\multicolumn{1}{l|}{}                             & \multicolumn{1}{l|}{Person-Appearance}    & The physical appearance of the person                                            \\ \cline{2-3} 
\multicolumn{1}{l|}{}                             & \multicolumn{1}{l|}{Person-Clothing}      & The clothing worn by the person                                                  \\ \cline{2-3} 
\multicolumn{1}{l|}{}                             & \multicolumn{1}{l|}{Person-Sex}           & The person's biological sex description                                          \\ \cline{2-3} 
\multicolumn{1}{l|}{}                             & \multicolumn{1}{l|}{Person-Action}        & An action or activity carried out by the person                                  \\ \cline{2-3} 
\multicolumn{1}{l|}{}                             & \multicolumn{1}{l|}{Person-Name}          & The name of the person                                                           \\ \cline{2-3} 
\multicolumn{1}{l|}{}                             & \multicolumn{1}{l|}{Person-Movement}      & The movement or change in location of the person                                 \\ \cline{2-3} 
\multicolumn{1}{l|}{}                             & \multicolumn{1}{l|}{Location-Description} & The descriptive information about the location                                   \\ \hline
\multicolumn{1}{l|}{\multirow{2}{*}{Contact}}     & \multicolumn{1}{l|}{PhoneNumber}          & A phone number                                                                   \\ \cline{2-3} 
\multicolumn{1}{l|}{}                             & \multicolumn{1}{l|}{Email}                & An email address                                                                 \\ \hline
\\ 
\hline \hline
\multicolumn{3}{c}{\textbf{Definition of Event Arguments}}                                                                                                                      \\ \hline \hline 
\multicolumn{2}{l|}{Argument}                                                                  & \textbf{Definition}                                                              \\ \hline
\multicolumn{2}{l|}{ATTACKER}                                                                 & The attacking/instigating agent                                                  \\ \hline
\multicolumn{2}{l|}{TARGET}                                                                   & The target of the offense (including unintended targets)                         \\ \hline
\multicolumn{2}{l|}{WEAPON}                                                                   & The weapon used in the offense                                                   \\ \hline
\multicolumn{2}{l|}{START-TIME}                                                               & When the incident starts                                                         \\ \hline
\multicolumn{2}{l|}{END-TIME}                                                                 & When the incident ends                                                           \\ \hline
\multicolumn{2}{l|}{PLACE}                                                                    & Where the incident takes place                                                   \\ \hline
\multicolumn{2}{l|}{TARGET-OBJECT}                                                            & The target object of the offense (e.g., vehicle, stolen items, etc.)             \\ \hline
\\ 
\hline \hline
\multicolumn{3}{c}{\textbf{Definition of Dispatchers' Intents}}                                                                                                                                                                                                        \\ \hline \hline
\multicolumn{2}{l|}{\textbf{Intent}}                                                          & \textbf{Definition}                                                                                                                                               \\ \hline
\multicolumn{2}{l|}{Thank}                                                                    & Expressing gratitude to the user                                                                                                                                  \\ \hline
\multicolumn{2}{l|}{ConfirmSendOfficer}                                                       & Confirming an officer has been dispatched                                                                                                                         \\ \hline
\multicolumn{2}{l|}{NotifyOthersInCharge}                                                     & Notifying other responsible personels/parties about the incident                                                                                                  \\ \hline
\multicolumn{2}{l|}{AskMeetOfficer}                                                           & Asking the user to meet an officer offline                                                                                                                        \\ \hline
\multicolumn{2}{l|}{AskToCall}                                                                & Asking the user to call an officer/the organization                                                                                                               \\ \hline
\multicolumn{2}{l|}{AskToVisit}                                                               & Asking the user to visit the organization's office                                                                                                                \\ \hline
\multicolumn{2}{l|}{AskForDetail}                                                             & \begin{tabular}[c]{@{}l@{}}Asking the user for additional detail related to the incident \\ (annotated with the event argument the dispatcher asks about, \\ i.e., ATTACKER, TARGET, WEAPON, START-TIME, \\ END-TIME, PLACE, TARGET-OBJECT)\end{tabular} \\ \hline
\end{tabular}
}
\end{table}

\clearpage
\newpage

\subsection{Examples of Dispatcher Responses Generated by the Chat Model}
\label{appdx:chat_example}

\begin{figure}[h!]
\resizebox{.95\textwidth}{!}{%
\begin{tabular}{p{.5\textwidth} p{.5\textwidth}}
\hline \hline
\textbf{Fine Tuning} & \textbf{RAG} \\ \hline \hline

\parbox[t]{.5\textwidth}{\small A chat between an individual reporting a safety concern to the local police department and a dispatcher from the police department. The dispatcher gives helpful and detailed guidance and instructions on how to proceed. The dispatcher is also supposed to give necessary emotional support to the user.} 
& \parbox[t]{.5\textwidth}{
\small A chat between an individual reporting a safety concern to the local police department and a dispatcher from the police department. The dispatcher gives helpful and detailed guidance and instructions on how to proceed. The dispatcher is also supposed to give necessary emotional support to the user. \\ \\ \\ 
\textbf{- Scenario: A user is reporting neighbors in the UV Suites for smoking weed indoors. The user lives on the terrace floor, and the neighbors, located in the \#\#\#\# building, are possibly in rooms \#\#\#\# \#\#\#\# or \#\#\#\#. The user declines the offer to meet with officers.\\Round 1: \\User: Hi. I live in the uv suites and I'm reporting my neighbors for smoking weed indoors. I live on the terrace floor where the trash chute is and the people smoking are all the way at {[}DATE{]}. The smell is strong and it's quite disturbing. \\Dispatcher: {[}ACRONYM{]}. Do you know the room number? Which building? \\Round 2 \\User: The \#\#\#\# building. It's either rooms \#\#\#\# \#\#\#\# or \#\#\#\# \\Dispatcher: Thank you so very much, I will send officers out. Would you like to meet with them? It's not required but I did want to offer. \\Round 3 \\User: No thanks\\} \\ 
}\\
\parbox[t]{.5\textwidth}{
\small - Scenario: A user reported a strong odor in a location, specifying it was in [GPE]. However, the user does not know the cause of the odor and expresses feeling very nervous about it. \\Round 3: \\User: Reeks of {[}GPE{]} in {[}GPE{]} \\Dispatcher: Do you have any descriptions on the cars?\\Round 2 \\User: I{[}ORG{]} not quite sure, I{[}ORG{]} sorry. I went to the bathroom and instantly smelled it in the hallway/bathroom. \\Dispatcher: \textcolor{bluec}{Which floor?}\\
} & 
\parbox[t]{.5\textwidth}{
\small - Scenario: A user reported a strong odor in a location, specifying it was in [GPE]. However, the user does not know the cause of the odor and expresses feeling very nervous about it. \\Round 1: \\User: Reeks of {[}GPE{]} in {[}GPE{]} \\Dispatcher: Hello, where at in {[}GPE{]} is the odor coming from? \\Round 2 \\User: I{[}ORG{]} not quite sure, I{[}ORG{]} sorry. I went to the bathroom and instantly smelled it in the hallway/bathroom. \\Dispatcher: \textcolor{bluec}{Which floor?}\\ 
}\\ \hline
\end{tabular}
}
\caption{
In the supervised fine-tuning, a prompted dialogue history along with the GPT-3.5 generated summary is aligned with the dispatcher's output. In evaluation, a similar document in the training set, based on the dialogue history along with its summary, is retrieved to promote a more human-like response from the dispatcher bot. 
}
    \label{fig:sft}
\end{figure}

\clearpage
\subsection{\yirenrevised{Survey Incident Case Samples}}
\label{apdx:survey_cases}

\begin{figure}[h!]
\resizebox{.95\textwidth}{!}{%
\begin{tabular}{p{.3\textwidth} p{.4\textwidth} p{.3\textwidth}}
\hline \hline
\textbf{User's Initial Message} &  & \\ \hline 

\parbox[t]{\textwidth}{\small [ORG] is in the [ORG] area of [ORG], loud music and talking. Can you send officers to check out? Thanks} 
& \\
\\
\hline
\hline
\textbf{Human Dispatcher} &  \textbf{GPT-3.5}&  \textbf{\textit{dispatcherLLM}}\\
\hline
\parbox[t]{.3\textwidth}{
\small Thank you for contacting [ORG]. Do you know how many people?\\
} & 
\parbox[t]{.4\textwidth}{
\small Hello, thank you for reporting this concern. I will definitely send officers to check out the situation. Can you please provide me with more details about the exact location and any other relevant information? It's important to stay calm and ensure your own safety while we handle this. Thank you for bringing this to our attention. We will do our best to address the situation promptly.\\ 
} & 
\parbox[t]{.3\textwidth}{
\small Hello, thank you for reporting this safety concern to us. Can you please provide me with more details about the loud music and talking that you are hearing?\\
}
\\ \hline
\end{tabular}
}
\caption{
A randomly selected \textit{Noise Disturbance} incident sample.
}
    \label{fig:survey_example_noise}
\end{figure}

\begin{figure}[h!]
\resizebox{.95\textwidth}{!}{%
\begin{tabular}{p{.3\textwidth} p{.4\textwidth} p{.3\textwidth}}
\hline \hline
\textbf{User's Initial Message} &  & \\ \hline 

\parbox[t]{\textwidth}{\small My wife was reading at the Nobel library in a single room this afternoon. A guy stormed into her room and grabbed her hand. She was on the \#nd floor. She is really scared and shocked by this incident. That guy was laughing while in her room. He has curly hair and was wearing a hoody. She couldn't give any other information regarding his appearance. Please kindly look into it. Thank you.} 
& \\
\\
\hline
\hline
\textbf{Human Dispatcher} &  \textbf{GPT-3.5}&  \textbf{\textit{dispatcherLLM}}\\
\hline
\parbox[t]{.3\textwidth}{
\small Thank you for contacting the [ACRONYM] Police. Do you know how long ago this happened?\\
} & 
\parbox[t]{.4\textwidth}{
\small I'm so sorry to hear about this terrifying experience. Thank you for reaching out to us. It's important that we take immediate action to ensure your wife's safety and apprehend the individual responsible. I want to make sure your wife is safe right now. Can you please confirm her current location and provide me with any other pertinent details? Additionally, I want you to know that you can stay on the line with me so that I can offer support and guidance as we address this situation together. Let's work together to keep your wife safe and ensure that this individual is held accountable for their actions.\\ 
} & 
\parbox[t]{.3\textwidth}{
\small Thank you for reporting this incident to us. I'm so sorry to hear that your wife had this frightening experience. I'm here to help and support you through this. Can you please tell me more about the location of the library? Is it a public library? And do you know the time of the incident?\\
}
\\ \hline
\end{tabular}
}
\caption{
A randomly selected \textit{Harassment / Abuse} incident sample.
}
    \label{fig:survey_example_harassment}
\end{figure}

\begin{figure}[h!]
\resizebox{.95\textwidth}{!}{%
\begin{tabular}{p{.3\textwidth} p{.4\textwidth} p{.3\textwidth}}
\hline \hline
\textbf{User's Initial Messages} &  & \\ \hline 

\parbox[t]{\textwidth}{\small User: [DATE] wasn't a good day. I've dealt with Depression before for [DATE] and there used to be many thoughts of ending it. As well as an "almost" attempt. I was doing fine and better, recently..[DATE]. It all just came out of nowhere and I'm just in a state of mind I don't want to be in. The thoughts came back [DATE]. \newline Dispatcher: Thank you for using LiveSafe. We would like to help. Where are you located or can you provide a phone number where we can have a counselor call you? \newline Dispatcher: Are you able to give us a call at (\#\#\#) \#\#\#-\#\#\#\#? \newline User: I would rather not talk over the phone if you don’t mind \newline User: I just finished my class at [ORG] library and was biking back home that’s why didn’t pick up. I apologize for that. \newline Dispatcher: That's understandable. I'm having an officer come up to text with you. \newline Dispatcher: Hello this is [PERSON]. [PERSON] \newline User: Hello how are you \newline User: Thank you \newline Dispatcher: I am concerned and would like to help you \newline Dispatcher: Are you willing to meetup? \newline Dispatcher: I would like to provide you with resources to help \newline User: I just got back home. And I would rather not in this weather. I’ve already hurt myself this morning and I wouldn’t want my bike to slip and I fall hurt myself again} 
& \\
\\
\hline
\hline
\textbf{Human Dispatcher} &  \textbf{GPT-3.5}&  \textbf{\textit{dispatcherLLM}}\\
\hline
\parbox[t]{.3\textwidth}{
\small Okay no problem\\
} & 
\parbox[t]{.4\textwidth}{
\small That's completely understandable, [PERSON]. I'm here to listen and support you in any way that I can. It's ok if you need some time to think about what you want to say. Just know that you're not alone in this, and there are people who care about you and want to help. Let's take things one step at a time.\\ 
} & 
\parbox[t]{.3\textwidth}{
\small I understand and I'm here for you. I want you to know that you are not alone and you are doing great.\\
}
\\ \hline
\end{tabular}
}
\caption{
A randomly selected \textit{Mental Health} incident sample.
}
    \label{fig:survey_example_mental_health}
\end{figure}

\clearpage
\subsection{\yirenrevised{Survey Questions}}
\label{apdx:survey_questions}
For each incident case sample, we ask the following survey questions:
\begin{itemize}
    \item Which model's suggestion do you think can be used by the human dispatcher to improve their response? (A, B, or Neither)
    \item If you select model A or B, please explain what part of the model's suggestion improves the human dispatcher's response.
    \item In the context of this (each) incident type, do you think the dispatcher should provide emotional support to the user?
    \item On a scale of 1 to 5, to what extent do you think Model A's response provides an appropriate level of emotional support?
    \item On a scale of 1 to 5, to what extent do you think Model B's response provides an appropriate level of emotional support?
\end{itemize}

After reviewing the incident samples, we ask each participant the following questions:
\begin{itemize}
    \item How do you think AI handles incidents with different levels of urgency (e.g., noise disturbance vs. gunshot)?
    \item Do you think AI should provide more emotional support for certain types of incidents? If so, what are the types of incidents?
\end{itemize}

\end{document}